\documentclass[12pt,a4paper]{article}

\usepackage{ifthen} 
\newboolean{pdflatex}
\setboolean{pdflatex}{true} 

\newboolean{articletitles}
\setboolean{articletitles}{true} 

\newboolean{uprightparticles}
\setboolean{uprightparticles}{false} 

\newboolean{inbibliography}
\setboolean{inbibliography}{false} 

\usepackage{graphpap} 
\usepackage{rotating} 
\usepackage{tikz}
\usepackage{xcolor}
\usepackage{rotating}
\usepackage{dashrule}
\usepackage{array} 
\usepackage{dcolumn}
\usepackage{comment}
\usepackage{xcolor}

\usepackage{longtable} 

\usepackage{lscape}
\usepackage{multirow} 
\usepackage{mathrsfs}
\usepackage{mathtools} 
\usetikzlibrary{patterns}
\usepackage{nicefrac} 
\usepackage{transparent}
\usepackage{afterpage} 

\def\paperauthors{LHCb collaboration} 
\def\paperasciititle{Study of the lineshape of the chi_c1(3872) state} 
\def\papertitle{Study of the lineshape 
of~the~$\chicone(3872)$ state} 
\def\paperkeywords{{High Energy Physics}, {LHCb}} 
\def\papercopyright{CERN on behalf of the LHCb collaboration}
\def\paperlicence{CC BY 4.0}
\def\paperlicenceurl{https://creativecommons.org/licenses/by/4.0/}

\makeatletter
\g@addto@macro\bfseries{\boldmath}
\makeatother

\usepackage[top=1in, bottom=1.25in, left=1in, right=1in]{geometry}

\usepackage{microtype}
\usepackage{lineno}  
\usepackage{xspace} 
\usepackage{caption} 

\usepackage{graphicx}  
\usepackage{color}
\usepackage{colortbl}
\graphicspath{{./figs/}} 

\usepackage{amsmath} 
\usepackage{amssymb}
\usepackage{amsfonts}
\usepackage{upgreek} 

\newcommand*\patchAmsMathEnvironmentForLineno[1]{%
\expandafter\let\csname old#1\expandafter\endcsname\csname #1\endcsname
\expandafter\let\csname oldend#1\expandafter\endcsname\csname
end#1\endcsname
 \renewenvironment{#1}%
   {\linenomath\csname old#1\endcsname}%
   {\csname oldend#1\endcsname\endlinenomath}%
}
\newcommand*\patchBothAmsMathEnvironmentsForLineno[1]{%
  \patchAmsMathEnvironmentForLineno{#1}%
  \patchAmsMathEnvironmentForLineno{#1*}%
}
\AtBeginDocument{%
\patchBothAmsMathEnvironmentsForLineno{equation}%
\patchBothAmsMathEnvironmentsForLineno{align}%
\patchBothAmsMathEnvironmentsForLineno{flalign}%
\patchBothAmsMathEnvironmentsForLineno{alignat}%
\patchBothAmsMathEnvironmentsForLineno{gather}%
\patchBothAmsMathEnvironmentsForLineno{multline}%
\patchBothAmsMathEnvironmentsForLineno{eqnarray}%
}

\usepackage{hyperxmp}

\usepackage[pdftex,
            pdfauthor={\paperauthors},
            pdftitle={\paperasciititle},
            pdfkeywords={\paperkeywords},
            pdfcopyright={Copyright (C) \papercopyright},
            pdflicenseurl={\paperlicenceurl}]{hyperref}

\usepackage[colorinlistoftodos,textsize=scriptsize]{todonotes}

\usepackage[bottom,flushmargin,hang,multiple]{footmisc}

\usepackage[all]{hypcap} 

\usepackage{xspace} 
\usepackage{upgreek}

\def\lhcb {\mbox{LHCb}\xspace}

\def\babar  {\mbox{BaBar}\xspace}
\def\belle  {\mbox{Belle}\xspace}

\def\MagUp {\mbox{\em Mag\kern -0.05em Up}\xspace}

\ifthenelse{\boolean{uprightparticles}}%
{

 \def\Pmu         {\ensuremath{\upmu}\xspace}

 \def\Ppi         {\ensuremath{\uppi}\xspace}                 
                  
 \def\Prho        {\ensuremath{\uprho}\xspace}

 \def\Pphi        {\ensuremath{\upphi}\xspace}                 
                  
 \def\Pchi        {\ensuremath{\upchi}\xspace}                 
 \def\Ppsi        {\ensuremath{\uppsi}\xspace}                 
 \def\Pomega      {\ensuremath{\upomega}\xspace}                 

 \def\PDelta      {\ensuremath{\Delta}\xspace}                 
 \def\PXi      {\ensuremath{\Xi}\xspace}                 
 \def\PLambda      {\ensuremath{\Lambda}\xspace}                 
 \def\PSigma      {\ensuremath{\Sigma}\xspace}                 
 \def\POmega      {\ensuremath{\Omega}\xspace}                 
 \def\PUpsilon      {\ensuremath{\Upsilon}\xspace}                 
 
 \def\PB      {\ensuremath{\mathrm{B}}\xspace}                 
                  
 \def\PD      {\ensuremath{\mathrm{D}}\xspace}

 \def\PJ      {\ensuremath{\mathrm{J}}\xspace}                 
 \def\PK      {\ensuremath{\mathrm{K}}\xspace}

 \def\PX      {\ensuremath{\mathrm{X}}\xspace}

 \def\Pb      {\ensuremath{\mathrm{b}}\xspace}                 
 \def\Pc      {\ensuremath{\mathrm{c}}\xspace}

 \def\Pi      {\ensuremath{\mathrm{i}}\xspace}

 \def\Ps      {\ensuremath{\mathrm{s}}\xspace}

}
{

 \def\Pmu         {\ensuremath{\mu}\xspace}

 \def\Ppi         {\ensuremath{\pi}\xspace}                 
                  
 \def\Prho        {\ensuremath{\rho}\xspace}

 \def\Pphi        {\ensuremath{\phi}\xspace}                 
                  
 \def\Pchi        {\ensuremath{\chi}\xspace}                 
 \def\Ppsi        {\ensuremath{\psi}\xspace}                 
 \def\Pomega      {\ensuremath{\omega}\xspace}                 
 \mathchardef\PDelta="7101
 \mathchardef\PXi="7104
 \mathchardef\PLambda="7103
 \mathchardef\PSigma="7106
 \mathchardef\POmega="710A
 \mathchardef\PUpsilon="7107
                  
 \def\PB      {\ensuremath{B}\xspace}                 
                  
 \def\PD      {\ensuremath{D}\xspace}

 \def\PJ      {\ensuremath{J}\xspace}                 
 \def\PK      {\ensuremath{K}\xspace}

 \def\PX      {\ensuremath{X}\xspace}

 \def\Pb      {\ensuremath{b}\xspace}                 
 \def\Pc      {\ensuremath{c}\xspace}

 \def\Pi      {\ensuremath{i}\xspace}

 \def\Ps      {\ensuremath{s}\xspace}

}

\makeatletter
\ifcase \@ptsize \relax
  \newcommand{\miniscule}{\@setfontsize\miniscule{4}{5}}
\or
  \newcommand{\miniscule}{\@setfontsize\miniscule{5}{6}}
\or
  \newcommand{\miniscule}{\@setfontsize\miniscule{5}{6}}
\fi
\makeatother

\DeclareRobustCommand{\optbar}[1]{\shortstack{{\miniscule (\rule[.5ex]{1.25em}{.18mm})}
  \\ [-.7ex] $#1$}}

\def\mup        {{\ensuremath{\Pmu^+}}\xspace}
\def\mun        {{\ensuremath{\Pmu^-}}\xspace} 

\def\squark    {{\ensuremath{\Ps}}\xspace}

\def\cquark    {{\ensuremath{\Pc}}\xspace}

\def\bquark    {{\ensuremath{\Pb}}\xspace}

\def\pion   {{\ensuremath{\Ppi}}\xspace}
\def\piz    {{\ensuremath{\pion^0}}\xspace}

\def\pip    {{\ensuremath{\pion^+}}\xspace}
\def\pim    {{\ensuremath{\pion^-}}\xspace}

\def\kaon    {{\ensuremath{\PK}}\xspace}
  \def\Kbar    {{\kern 0.2em\overline{\kern -0.2em \PK}{}}\xspace}

\def\KorKbar    {\kern 0.18em\optbar{\kern -0.18em K}{}\xspace}

\def\Kp      {{\ensuremath{\kaon^+}}\xspace}

\def\KS      {{\ensuremath{\kaon^0_{\mathrm{S}}}}\xspace}

  \def\Dbar    {{\kern 0.2em\overline{\kern -0.2em \PD}{}}\xspace}
\def\D       {{\ensuremath{\PD}}\xspace}

\def\DorDbar    {\kern 0.18em\optbar{\kern -0.18em D}{}\xspace}
\def\Dz      {{\ensuremath{\D^0}}\xspace}
\def\Dzb     {{\ensuremath{\Dbar{}^0}}\xspace}
\def\Dp      {{\ensuremath{\D^+}}\xspace}

\def\Dstarb  {{\ensuremath{\Dbar{}^*}}\xspace}
\def\Dstarz  {{\ensuremath{\D^{*0}}}\xspace}
\def\Dstarzb {{\ensuremath{\Dbar{}^{*0}}}\xspace}
\def\Dstarp  {{\ensuremath{\D^{*+}}}\xspace}
\def\Dstarm  {{\ensuremath{\D^{*-}}}\xspace}

\def\B       {{\ensuremath{\PB}}\xspace}
\def\Bbar    {{\ensuremath{\kern 0.18em\overline{\kern -0.18em \PB}{}}}\xspace}

\def\BorBbar    {\kern 0.18em\optbar{\kern -0.18em B}{}\xspace}
\def\Bz      {{\ensuremath{\B^0}}\xspace}

\def\Bu      {{\ensuremath{\B^+}}\xspace}

\def\Bs      {{\ensuremath{\B^0_\squark}}\xspace}

\def\jpsi     {{\ensuremath{{\PJ\mskip -3mu/\mskip -2mu\Ppsi\mskip 2mu}}}\xspace}
\def\psitwos  {{\ensuremath{\Ppsi{(2S)}}}\xspace}

\def\chicone  {{\ensuremath{\Pchi_{\cquark 1}}}\xspace}

  \def\Y#1S{\ensuremath{\PUpsilon{(#1S)}}\xspace}

\def\Lbar        {{\ensuremath{\kern 0.1em\overline{\kern -0.1em\PLambda}}}\xspace}
\def\LorLbar    {\kern 0.18em\optbar{\kern -0.18em \PLambda}{}\xspace}

\newcommand{\decay}[2]{\ensuremath{#1\!\to #2}\xspace}         

\def\to                 {\ensuremath{\rightarrow}\xspace}

\def\AT#1     {\ensuremath{A_{\mathrm{T}}^{#1}}\xspace}           

\def\C#1      {\ensuremath{\mathcal{C}_{#1}}\xspace}                       
\def\Cp#1     {\ensuremath{\mathcal{C}_{#1}^{'}}\xspace}                    
\def\Ceff#1   {\ensuremath{\mathcal{C}_{#1}^{\mathrm{(eff)}}}\xspace}        
\def\Cpeff#1  {\ensuremath{\mathcal{C}_{#1}^{'\mathrm{(eff)}}}\xspace}       
\def\Ope#1    {\ensuremath{\mathcal{O}_{#1}}\xspace}                       
\def\Opep#1   {\ensuremath{\mathcal{O}_{#1}^{'}}\xspace}                    

\newcommand{\tev}{\ifthenelse{\boolean{inbibliography}}{\ensuremath{~T\kern -0.05em eV}}{\ensuremath{\mathrm{\,Te\kern -0.1em V}}}\xspace}
\newcommand{\gev}{\ensuremath{\mathrm{\,Ge\kern -0.1em V}}\xspace}
\newcommand{\mev}{\ensuremath{\mathrm{\,Me\kern -0.1em V}}\xspace}
\newcommand{\kev}{\ensuremath{\mathrm{\,ke\kern -0.1em V}}\xspace}
\newcommand{\ev}{\ensuremath{\mathrm{\,e\kern -0.1em V}}\xspace}
\newcommand{\gevc}{\ensuremath{{\mathrm{\,Ge\kern -0.1em V\!/}c}}\xspace}
\newcommand{\mevc}{\ensuremath{{\mathrm{\,Me\kern -0.1em V\!/}c}}\xspace}
\newcommand{\gevcc}{\ensuremath{{\mathrm{\,Ge\kern -0.1em V\!/}c^2}}\xspace}
\newcommand{\gevgevcccc}{\ensuremath{{\mathrm{\,Ge\kern -0.1em V^2\!/}c^4}}\xspace}
\newcommand{\mevcc}{\ensuremath{{\mathrm{\,Me\kern -0.1em V\!/}c^2}}\xspace}

\def\mum  {\ensuremath{{\,\upmu\mathrm{m}}}\xspace}

\def\invfb   {\ensuremath{\mbox{\,fb}^{-1}}\xspace}

\newcommand{\chisq}{\ensuremath{\chi^2}\xspace}

\newcommand{\chisqip}{\ensuremath{\chi^2_{\text{IP}}}\xspace}

\def\gsim{{~\raise.15em\hbox{$>$}\kern-.85em
          \lower.35em\hbox{$\sim$}~}\xspace}
\def\lsim{{~\raise.15em\hbox{$<$}\kern-.85em
          \lower.35em\hbox{$\sim$}~}\xspace}

\def\ptot       {\mbox{$p$}\xspace}
\def\pt         {\mbox{$p_{\mathrm{ T}}$}\xspace}

\def\evtgen     {\mbox{\textsc{EvtGen}}\xspace}

\def\geant      {\mbox{\textsc{Geant4}}\xspace}

\def\photos     {\mbox{\textsc{Photos}}\xspace}

\def\pythia     {\mbox{\textsc{Pythia}}\xspace}

\def\tell1  {TELL1\xspace}
\def\ukl1   {UKL1\xspace}

\newcommand{\eg}{\mbox{\itshape e.g.}\xspace}

\usepackage{cite} 
\usepackage{mciteplus}

\usepackage{longtable} 

\begin{document}

\renewcommand{\thefootnote}{\fnsymbol{footnote}}
\setcounter{footnote}{1}


\begin{titlepage}
\pagenumbering{roman}

\vspace*{-1.5cm}
\centerline{\large EUROPEAN ORGANIZATION FOR NUCLEAR RESEARCH (CERN)}
\vspace*{1.5cm}
\noindent
\begin{tabular*}{\linewidth}{lc@{\extracolsep{\fill}}r@{\extracolsep{0pt}}}
\ifthenelse{\boolean{pdflatex}}
{\vspace*{-1.5cm}\mbox{\!\!\!\includegraphics[width=.14\textwidth]{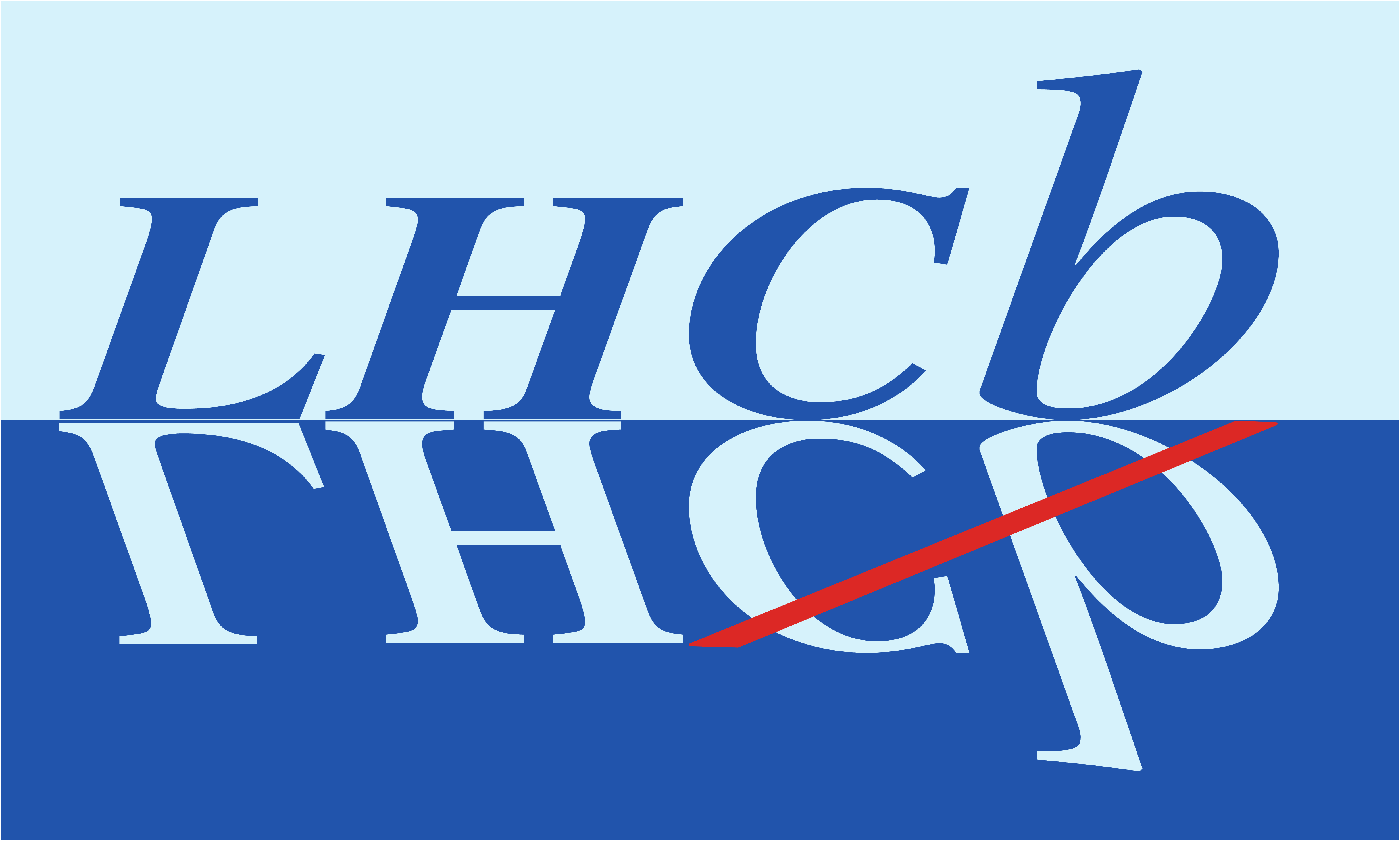}} & &}%
{\vspace*{-1.2cm}\mbox{\!\!\!\includegraphics[width=.12\textwidth]{figs/lhcb-logo.eps}} & &}%
\\
 & & CERN-EP-2020-086\\  
 & & LHCb-PAPER-2020-008 \\  
 & & March 11, 2021 \\ 
\end{tabular*}

\vspace*{2.0cm}

{\normalfont\bfseries\boldmath\huge
\begin{center}
  \papertitle
\end{center}
}

\vspace*{2.0cm}

\begin{center}
\paperauthors\footnote{Authors are listed at the end of this paper.}
\end{center}

\vspace{\fill}

\begin{abstract}
\noindent  A study of the lineshape of the $\chicone(3872)$ state is
made using a data sample corresponding to an integrated
luminosity of $3\invfb$ collected in $pp$ collisions at
centre-of-mass energies of 7 and 8\,TeV with the LHCb
detector. Candidate $\chicone(3872)$ and $\psitwos$ mesons from $\bquark$-hadron
decays are selected in the $ J/\psi \pi^+ \pi^-$ decay mode.
Describing the {\mbox{lineshape}} with a Breit--Wigner function,
the mass splitting between the
$\chicone(3872)$ and $\psitwos$ states, $\Delta m$,
and the width of the $\chicone(3872)$ state,
$\Gamma_{\mathrm{BW}}$,
are determined to be
\begin{eqnarray*}
\Delta m & = & 185.598
\pm 0.067  \pm 0.068 \mev \,, \\
\Gamma_{\mathrm{BW}} & = &  \phantom{00}1.39\phantom{0}
\pm 0.24\phantom{0} \pm 0.10\phantom{0} \mev  \,,
\end{eqnarray*}
where the first uncertainty is statistical and the second  systematic.
Using a Flatt\'e-inspired model, the mode and full width at half maximum of the lineshape are determined to be 
\begin{eqnarray*}
\mathrm{mode} & =  3871.69^{\,+\,0.00\,+\,0.05}_{-0.04\,-\,0.13} &\mev \\
\mathrm{FWHM} & =  0.22^{\,+\,0.07\,+\,0.11}_{\,-\,0.06\,-\,0.13}& \mev .
\end{eqnarray*}
An investigation of the  analytic structure of the Flatt\'e amplitude reveals a pole structure, which is compatible
with a quasi-bound $\Dz\Dstarzb$~state but a quasi-virtual state is still allowed
at the level of $2$~standard deviations.
\end{abstract}

\vspace*{1.0cm}

\begin{center}
 Published in Phys.~Rev.~D102 (2020) 092005
\end{center}

\vspace{\fill}

{\footnotesize
\centerline{\copyright~\papercopyright, licence \href{\paperlicenceurl}{\paperlicence}.}}

\end{titlepage}


\newpage
\setcounter{page}{2}
\mbox{~}
%

\cleardoublepage


\renewcommand{\thefootnote}{\arabic{footnote}}
\setcounter{footnote}{0}



\pagestyle{plain} 
\setcounter{page}{1}
\pagenumbering{arabic}


%


\section{Introduction}
\label{sec:Introduction}
The last two decades have seen a resurgence of interest in the spectroscopy of
non\nobreakdash-conventional\,(exotic) charmonium states~\cite{swanson}
starting with the observation of the
charmonium-like $\chicone(3872)$ state by
the Belle collaboration~\cite{Choi:2003ue}.
Though the existence of the $\chicone(3872)$~particle has
been confirmed by many experiments~\cite{Ablikim:2013dyn,
  Aaltonen:2009vj, Aubert:2008g, Abazov:2004kp,LHCb-PAPER-2011-034}
with quantum numbers measured
to be $1^{++}$~\cite{LHCb-PAPER-2013-001,LHCb-PAPER-2015-015},
its nature is still uncertain. Several exotic
interpretations have been suggested: \eg a
tetraquark~\cite{Maiani:2004vq}, a loosely bound deuteron\nobreakdash-like
 $\Dz\Dstarzb$~molecule~\cite{Tornqvist:2004qy} or a
 charmonium\nobreakdash-molecule mixture~\cite{Hanhart:2011jz}.

A striking feature of the $\chicone(3872)$~state is
the proximity of its mass
to the sum of the $\Dstarz$ and $\Dz$ meson
masses.
Accounting for correlated uncertainties due to
the knowledge of the kaon mass,
this sum is evaluated to be
\mbox{$m_{\Dz} + m_{\Dstarz} = 3871.70 \pm  0.11 \mev$} \footnote{
Natural units with $c=\hbar=1$ are used through this paper.
}.
The molecular interpretation of
the $\chicone(3872)$ state requires it to be a bound state.
Assuming a Breit--Wigner lineshape, this implies
that
$\delta E \equiv m_{\Dz} + m_{\Dstarz} - m_{\chicone(3872)} > 0$.
Current knowledge of
$\delta E$
is limited by the
uncertainty on the $\chicone(3872)$ mass, motivating a more precise
determination of this quantity. The nature of the $\chicone(3872)$~state can
also be elucidated by studies of its lineshape.
This has been
analysed by several experiments
assuming a Breit--Wigner
function~\cite{Ablikim:2013dyn,Aubert:2008g, Aubert:2005zh}. The current
upper limit on the  natural width,
$\Gamma_{\mathrm{BW}}$, is \mbox{$1.2 \mev $}
at $90 \%$ confidence level~\cite{Choi:2011fc}.

In this analysis a sample of
$\chicone(3872) \rightarrow J/\psi \pi^+ \pi^-$ candidates
produced in inclusive $\bquark$-hadron decays is used
to measure precisely the mass and to determine  the
lineshape of the $\chicone(3872)$ meson. Studies are made
assuming both a Breit--Wigner
lineshape and  a Flatt\'e-inspired model that accounts
for the opening up of the
$\Dzb \Dstarz$ threshold~\cite{PhysRevD.76.034007,Kalashnikova:2009gt}.
The analysis uses a data sample  corresponding to an
integrated luminosity of $3\invfb$ of data collected in $pp$ collisions at
centre-of-mass energies of 7~TeV and 8~TeV during 2011 and 2012 using
the LHCb detector.


\section{Detector and simulation}
\label{sec:Detector}
The \lhcb detector~\cite{Alves:2008zz,LHCb-DP-2014-002} is a
single-arm forward spectrometer covering the \mbox{pseudorapidity}
range \mbox{$2<\eta <5$}, designed for the study of particles containing
\bquark or \cquark quarks. The detector includes a high-precision
tracking system consisting of a silicon-strip vertex detector
surrounding the $pp$ interaction
region~\cite{LHCb-DP-2014-001}, a large-area silicon-strip
detector (TT) located upstream of a dipole magnet with a bending power of
about $4{\mathrm{\,Tm}}$, and three stations of silicon-strip
detectors and straw drift tubes~\cite{LHCb-DP-2013-003} placed
downstream of the magnet. The tracking system provides a measurement
of momentum, \ptot, of charged particles with a relative uncertainty
that varies from 0.5\% at low momentum to 1.0\% at 200\gev. As
described in Refs.~\cite{LHCb-PAPER-2012-048,LHCb-PAPER-2013-011} the
momentum scale is calibrated using samples of $\jpsi \rightarrow \mup
\mun$ and $\Bu \rightarrow \jpsi \Kp$ decays collected concurrently
with the data sample used for this analysis. The~relative accuracy of this
procedure is estimated to be $3 \times 10^{-4}$ using samples of other
fully reconstructed $\bquark$~hadrons, $\PUpsilon$~and
$\KS$~mesons. The minimum distance of a track to a primary vertex\,(PV), the impact
parameter\,(IP), is measured with a resolution of $(15+29/\pt)\mum$,
where \pt is the component of the momentum transverse to the beam,
in\,\gev.

Different types of charged
hadrons are distinguished using information from two
ring\nobreakdash-imaging Cherenkov\,(RICH)
detectors. Photons, electrons and hadrons are
identified by a calorimeter system consisting of scintillating-pad and
preshower detectors, an electromagnetic calorimeter and a hadronic
calorimeter. Muons are  identified by a system composed of alternating
layers of iron and multiwire proportional chambers~\cite{LHCb-DP-2012-002}.

The online event selection is performed by a
trigger~\cite{LHCb-DP-2012-004}, which consists of a hardware stage,
based on information from the calorimeter and muon
systems, followed by a software stage, where  a full event
reconstruction is made. Candidate events are required to pass the
hardware trigger, which selects muon and dimuon candidates with high
$\pt$ based upon muon system information. The subsequent software trigger
is composed of two stages. The first performs a
partial event reconstruction and requires events to have two well-identified
oppositely charged muons and that the mass of the pair is larger
than $2.7\gev$. The second stage performs a full event
reconstruction.  Events are retained for further processing if they
contain a displaced $\mup\mun$ vertex. The
decay vertex is required to be well separated from each reconstructed
PV of the proton-proton interaction by requiring the distance between
the PV and the $\mup\mun$ vertex  divided by its uncertainty to be greater than three.

To study the properties of the signal and the most important backgrounds,
simulated sampels of $pp$ collisions are generated using
\pythia~\cite{Sjostrand:2006za,*Sjostrand:2007gs}  with a specific
\lhcb configuration~\cite{LHCb-PROC-2010-056}.  Decays of hadronic
particles are described by \evtgen~\cite{Lange:2001uf}, in which
final-state radiation is generated using
\photos~\cite{Golonka:2005pn}. The
interaction of the generated particles with the detector, and its
response, are implemented using the \geant
toolkit~\cite{Allison:2006ve, *Agostinelli:2002hh} as described in
Ref.~\cite{LHCb-PROC-2011-006}. For the study of the lineshape it is
important that the simulation models well the mass resolution.
The simulation used in this study reproduces the observed mass resolution for selected samples of
$\Bu \rightarrow \jpsi \Kp$,
$\Bz \rightarrow \jpsi \Kp \pim$,
$\Bs \rightarrow \jpsi \Pphi$ and
$\Bu \rightarrow \jpsi  \Kp \pip \pim$ decays within $5 \%$.
To further improve the agreement for the mass resolution between data and simulation,
scale factors are determined using a large sample of
$\psitwos \rightarrow \jpsi
\pip \pim $ decays collected concurrently with the
$\chicone(3872)$ sample. This will be discussed in detail below. 


\section{Selection}
\label{sec:selection}
The selection of $\chicone(3872) \rightarrow \jpsi \pip \pim$ candidates
from $\bquark$-hadron decays is performed in two steps.
First, loose selection criteria are applied that
reduce the background from random combinations of tracks
significantly while retaining
high signal efficiency. Subsequently, a
multivariate selection is used to further reduce this combinatorial
background. In both steps, the selection requirements are chosen to reduce background
whilst selecting well reconstructed candidates.
The requirements are optimised using simulated signal decays together with
a sample of selected candidates in data where
the charged pions have the same sign. The latter sample is found to be
a good proxy to describe the background shape.
Though the selection criteria are tuned using the $\chicone(3872)$ simulation
sample,  the $\psitwos \rightarrow \jpsi \pi^+ \pi^-$ decay mode is also
selected with high efficiency and used for calibration.

The selection starts from a pair of oppositely charged particles,
identified as muons. Incorrectly reconstructed tracks
are suppressed by imposing a requirement on the output
of a neural network trained to discriminate between these and trajectories from real
particles. To select $\jpsi \rightarrow \mup \mun$ candidates, the
two muons are required to originate from a common vertex that is
significantly displaced from any PV. The difference between the reconstructed
invariant mass of the pair and the
known value of the $\jpsi$ mass~\cite{PDG2019} is required to be within three
times the uncertainty on the reconstructed mass of the $\mup\mun$~pair.

Pion candidates are selected
using the same track-quality requirements as the
muons. Information from the muon system is used to reject pions
that decayed in the spectrometer since these pions
tend to have poorly reconstructed trajectories which result in
$\chicone(3872)$ candidates with worse mass resolution.
Combinatorial
background is suppressed by requiring that the \chisqip
of the pion candidates, defined as the difference between
the \chisq of the PV reconstructed with and without
the considered particle, is larger than
four for all PVs.
Good pion identification is ensured
by applying a requirement on a variable that combines
information from the RICH detectors with kinematic
and track quality information.
Since the pions produced in
$\chicone(3872)$ decays have relatively small \pt,
only a loose requirement on
the transverse momentum\,($\pt > 200 \mev$) is imposed.
In addition, the pion candidates are required to have
\mbox{$p <50 \gev$}. This requirement rejects
candidates with poor momentum resolution
and has an efficiency of $99.5 \, \%$.

To create $\chicone(3872)$ candidates,
$J/\psi$ candidates are combined
with pairs of oppositely charged pions.
To improve the
mass resolution
a kinematic vertex fit
\cite{Hulsbergen:2005pu} is made which constrains the $J/\psi$ invariant mass to
its known value~\cite{PDG2019}.
The
reduced $\chi^2$~of the fit, $\chi^2_{\mathrm{fit}}/\mathrm{ndf}$,
is required to be less than five.
Candidates
with a mass uncertainty greater than $5.0 \mev$ are rejected.
Finally, requiring the $Q$\nobreakdash-value
of the decay to be below 200\mev
substantially reduces the
background whilst retaining $96 \%$ of the $\chicone(3872)$
signal. Here the $Q$\nobreakdash-value  is defined as
\mbox{$Q \equiv
m_{\mup\mun\pip\pim} - m_{\mup\mun} - m_{\pip\pim}$} where $m_{\mup\mun\pip\pim}$,
$m_{\mup\mun}$ and
$m_{\pip\pim}$
are the reconstructed masses of the final state combinations.

The final step of the selection process is based on a neural network
classifier~\cite{McCulloch,
rosenblatt58,Zhong:2011xm,
Hocker:2007ht,*TMVA4}. This is trained on a simulated sample
of inclusive $\bquark \rightarrow \chicone(3872)\PX$
decays and the same-sign pion
sample in data.
Simulated samples are corrected
to reproduce kinematical distributions
of the \psitwos mesons observed on data.
The training is performed separately for the 2011 and
2012 data samples.
Twelve variables that give good separation between signal and
background are
considered:
the pseudorapidity and transverse momentum
of the two pion candidates, the \chisqip for each of the two pions,
the pseudorapidity and transverse momentum of
the $\chicone(3872)$ candidate, the $\chisq$ of
the two-track vertex fit for the pions,
the $\chi^2_{\mathrm{fit}}/\mathrm{ndf}$,
the flight distance $\chi^2$ of
the candidate calculated using the reconstructed
primary and secondary vertices, and the total
number of hits in the TT detector.  All these variables show good agreement between
simulation and data. The optimal cut on the classifier output is chosen
using pseudoexperiments
so as to minimise the uncertainty on the measured $\chicone(3872)$ mass.


\section{Mass model}
\label{sec:model}
The observed invariant mass distribution of the $\jpsi\pip\pim$ system,
$m_{\jpsi\pip\pim}$,  for the
$\psitwos$ and $\chicone(3872)$ resonances is a
convolution of the
natural lineshape with the detector resolution. For the $\psitwos$
resonance the lineshape is well described by a Breit--Wigner function.
The situation for the $\chicone(3872)$ meson is more
complex. Previous measurements have assumed
a Breit--Wigner resonance shape.
However, as discussed in Refs.~\cite{PhysRevD.76.034007,Kalashnikova:2009gt,Hanhart:2011jz},
this is not well motivated due to the proximity
of the $\Dz\Dstarzb$~threshold. Several
 other alternative lineshapes have been proposed in the
 literature~\cite{PhysRevD.76.034007,Kalashnikova:2009gt, Braaten:2007dw,Stapleton:2009ey}.
 In this analysis two lineshapes
for the $\chicone(3872)$ meson are considered in detail, a Breit--Wigner
and a Flatt\'e-inspired model~\cite{PhysRevD.76.034007,Kalashnikova:2009gt}.
These models are investigated in the next sections.
The S-wave threshold resonance  model described in Ref.~\cite{Braaten:2007dw,Stapleton:2009ey},
that accounts
for the non-zero width of the $D^{*0}$ meson, was considered but did not fit the data well.
If the mass is close to the $\Dzb \Dstarz$ threshold, this model is not able to
accommodate a value of the natural width much larger than
\mbox{$\Gamma_{\Dstarz} = 65.5 \pm 15.4 \kev$}~\cite{Braaten:2007dw}.
As  will be discussed below, the study presented
here favours larger values of the natural width.


\label{sec:resolution}
The analysis proceeds in two steps. First,
unbinned maximum\nobreakdash-likelihood
fits are made to the $m_{\jpsi\pip\pim}$
distribution in the region around the $\psitwos$ mass.
These measured values of the $\psitwos$ mass and mass resolution
are used to control systematic uncertainties in the subsequent fits
to the $m_{\jpsi\pip\pim}$ distribution
in the~$\chicone(3872)$ mass region.
For both sets of fits the natural lineshape is convolved with a
resolution model developed using the simulation.
The application of the $\jpsi$ mass constraint
in the fit~\cite{Hulsbergen:2005pu} results in the mass resolution
being dominated by the kinematics of the pion pair. In particular, the
resolution is worse for higher values of the total momentum of the
pion pair, $p_{\pip\pim}$.
Consequently,
the analysis is performed
in three $p_{\pip\pim}$~bins chosen
to contain an approximately equal number of signal candidates:
\mbox{$p_{\pip\pim} < 12 \gev$},
\mbox{$ 12 \le p_{\pip \pim} < 20 \gev  $} and
\mbox{$20 \le  p_{\pip \pim} < 50  \gev $}.
The core mass resolution for the $\chicone(3872)$ state varies
monotonically between $2.4 \mev$ and $3.0 \mev$ between
the lowest\nobreakdash-$p_{\pip\pim}$ and highest\nobreakdash-$p_{\pip\pim}$
bin. Possible differences in data taking conditions are allowed for by
dividing the data  according to the year of collection
resulting in a total of six data samples.

The resolution model is studied using simulation.
In each $p_{\pip\pim}$ bin the mass resolution  is modelled
with the sum of a narrow Crystal Ball
function~\cite{Skwarnicki:1986xj} combined with a wider Gaussian
function.  The Crystal Ball function has a
Gaussian core and two parameters that describe
the power-law tail.
The simulation is also used to determine the value of
the transition point between the core and the power law tail,
$a$, as a multiple of the width, $\sigma$, of the
Gaussian core. The value of the exponent of the power law, $n$,
is allowed to vary
in the data fits  with a Gaussian constraint to the value obtained
in the simulation applied.  When fitting the $\chicone(3872)$ mass
region in data the values of the core resolution, $\sigma$, for the
Gaussian and Crystal Ball functions are
taken from simulation up to an overall scale factor,
$s_f$, that
accounts for residual discrepancies between data and simulation.
For each $p_{\pip\pim}$ data sample
the value of $s_f$ is determined  in the corresponding fit to
the $\psitwos$ mass region and applied as a Gaussian
constraint. The systematic uncertainty associated
with the choice of the signal model is assessed
by replacing the nominal model with the sum of either
two Crystal Ball or Gaussian functions.

The shape of the combinatorial background is studied using the
same-sign data sample as well as samples of simulated
inclusive $b \rightarrow J/\psi\PX$ decays.
Based upon these studies,
the background is modelled by
the form
\mbox{$(m_{\jpsi\pip\pim}-m_{\jpsi}-2m_{\pi^{\pm}})^{c_0}\, e^{-m_{\jpsi\pip\pim}/c_1}$},
where
$c_0$ is fixed to 3.6 based on fits to the
same-sign data. Variations of this functional form together with other
models (\eg exponential or polynomial functions) are
used as systematic variations. In total, seven different background
forms are considered.


\section{\psitwos mass}
\label{sec:psifits}
\begin{figure*}[t]
\includegraphics*[width=0.8\textwidth]{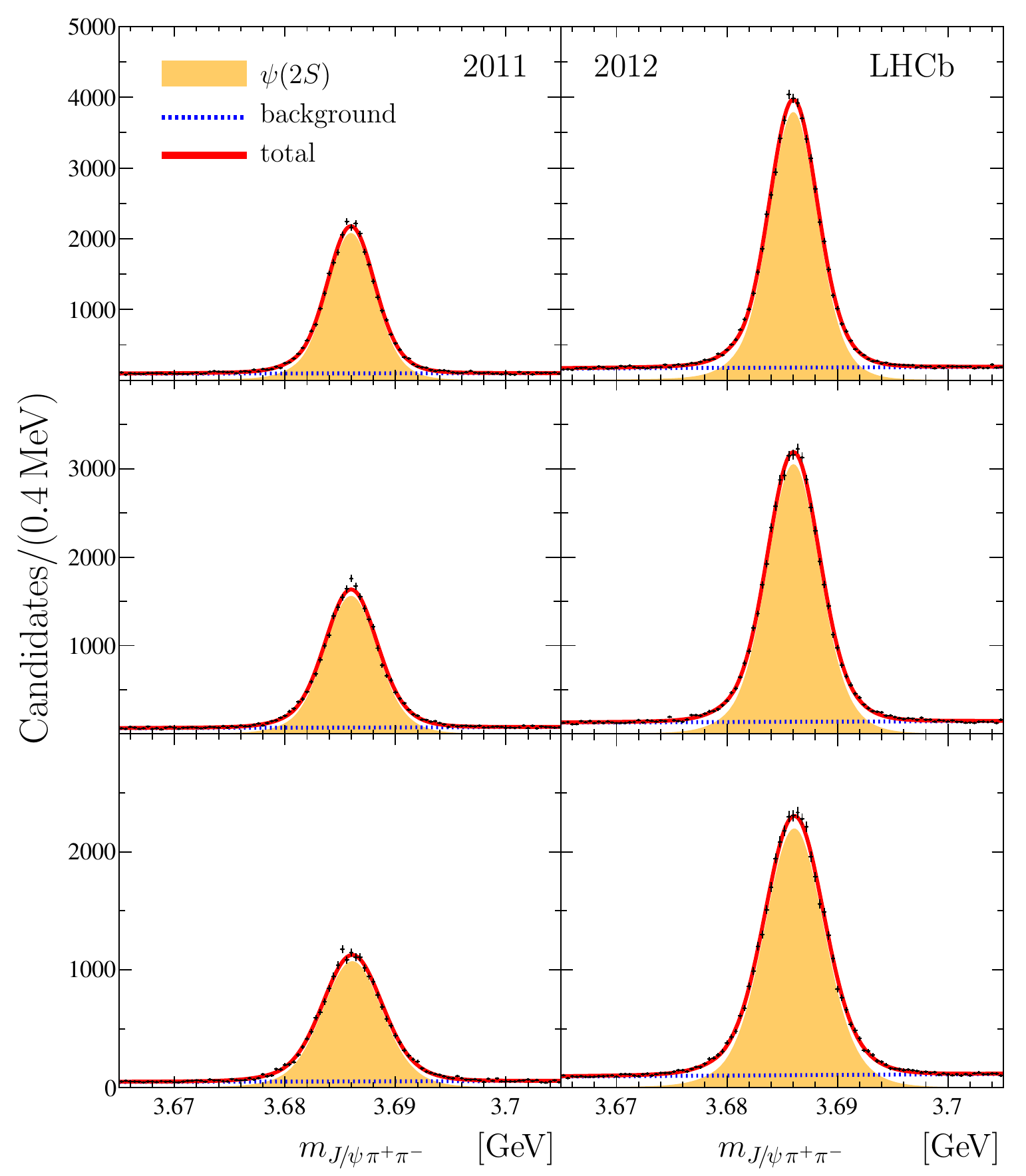}
\caption{\small
  Mass distributions for $\jpsi\pip\pim$~candidates
  in the $\psitwos$ region for (top)\,the low,
  (middle)\,mid and (bottom)\,high $p_{\pip\pim}$~bins.
  The left\,(right)-hand plot is for 2011\,(2012) data.
  The
  projection of the fit described in the text is
  superimposed.
  }
\label{fig:psipolx}
\end{figure*}

Since the \psitwos state is narrow and away from the phase-space limits,
a spin-0 relativistic Breit-Wigner function is used
to model the lineshape.
A spin-1 Breit-Wigner function is considered as
part of the systematic uncertainties and found to give identical results.
This lineshape is convolved
with the default resolution model 
and a fit to $\jpsi \pip \pim$ mass is performed
in each of the six $p_{\pip\pim}$ data samples.
The natural width of the $\psitwos$ is fixed
to the known value \cite{PDG2019}. Figure~\ref{fig:psipolx} shows
the $m_{\jpsi \pip \pim}$ distributions and
fit projections for each data sample and Table~\ref{tab:psicalibs}
summarises the resulting parameters of
interest.
Binning the data and calculating the \chisq~probability
of consistency with the fit model
gives the values  greater than 5\%
for all fits.
The fitted values of
the \psitwos mass agree with the known value~\cite{PDG2019}
within the uncertainty of the calibration procedure.
The values of $s_f$
are consistent with the expectation that
the simulation reproduces the
mass resolution in data at the level
of $5 \%$ or better.
When applied as constraint in the fit to the $\chicone(3872)$ region, additional
uncertainties on $s_f$ are considered. Accounting for the finite size
of the simulation samples,
the background modelling and the assumption
that the $\psi(2S)$
calibration factor can be applied to the $\chicone(3872)$
candidates,
the uncertainty on $s_f$ is 0.02, independent of the bin.
The values of $s_f$
in Table \ref{tab:psicalibs} are applied as
Gaussian constraints in the
fits to the $\chicone(3872)$ region with an uncertainty of 0.02.

\begin{table}[b]
  \centering
  \caption{\small Results of the~$\psitwos$ mass and
  scale factor  $s_f$ obtained for the
    nominal fit model. The quoted uncertainties on the
    $\psitwos$ mass and $s_f$ are statistical.}  \label{tab:psicalibs}
  \vspace*{3mm}
	\begin{tabular*}{\columnwidth}{@{\hspace{3mm}}r@{\extracolsep{\fill}}rcc@{\hspace{3mm}}}
	Year & $p_{\pip\pim}~\left[\!\gev\right]$\ \
	& $m_{\psitwos}~\left[\!\mev\right]$  & $s_f$
    \\[1.5mm]
    \hline
    \\[-3mm]
    2011 &  $\phantom{00\le}p_{\pip\pim}<12$
    & $ 3685.97 \pm 0.02 $ &  $ 1.03 \pm 0.01 $
    \\
    2011 &  $12\le p_{\pip\pim}<20$
    & $  3685.98 \pm 0.02 $ &  $ 1.05 \pm  0.01$
    \\
    2011 & $20\le p_{\pip\pim}<50$
   & $ 3686.10 \pm 0.03 $ &  $ 1.04 \pm 0.01 $
   \\
    2012 & $\phantom{00\le}p_{\pip\pim}<12$
       & $ 3686.01 \pm 0.01 $ &  $ 1.03 \pm 0.01 $
    \\
    2012 & $12\le p_{\pip\pim}<20$
     & $ 3686.02\pm 0.01 $ &  $ 1.05 \pm 0.01 $
    \\
     2012 & $20\le p_{\pip\pim}<50$
    & $ 3686.09 \pm 0.02 $ &  $ 1.01 \pm 0.01  $
 \end{tabular*}
\end{table}


\section{Breit--Wigner mass and width  of the $\chicone(3872)$ state}
\label{sec:bw}
To extract the Breit--Wigner lineshape parameters of the
$\chicone(3872)$ meson, a fit is made to the mass range
$3832 < m_{\jpsi \pip\pim} < 3912\mev$ in each of
the six $p_{\pip\pim}$ data samples
described above. 
A spin-0 relativistic Breit-Wigner is used, as in Ref.~\cite{LHCb-PAPER-2015-015},

For each data sample the mass difference
between the $\psitwos$ and $\chicone(3872)$ meson,
$\Delta m$, is measured relative to the measured
mass of the \psitwos~state rather than the
absolute mass. This minimises the systematic
uncertainty due to the momentum scale.
The fit in each bin has seven free parameters:
$\Delta m$, the natural
width $\Gamma_{\mathrm{BW}}$,
the background parameter $c_1$,
the resolution scale factor $s_f$,
the tail parameter $n$,  and the signal  and background
yields. Again 
a Gaussian constraint is applied to $n$ based on the simulation. The parameter
$s_f$ is constrained to the result of the fit to the $\psitwos$ data. 
The fit procedure is validated using
both the simulation and pseudoexperiments. No significant bias is
found and the uncertainties estimated by the fit agree with
the spread observed
in the pseudoexperiments. These studies show that,
values of $\Gamma_{\mathrm{BW}}$ larger than $0.6 \mev$ can
reliably be determined.

\begin{figure*}[t]
\includegraphics*[width=0.8\textwidth]{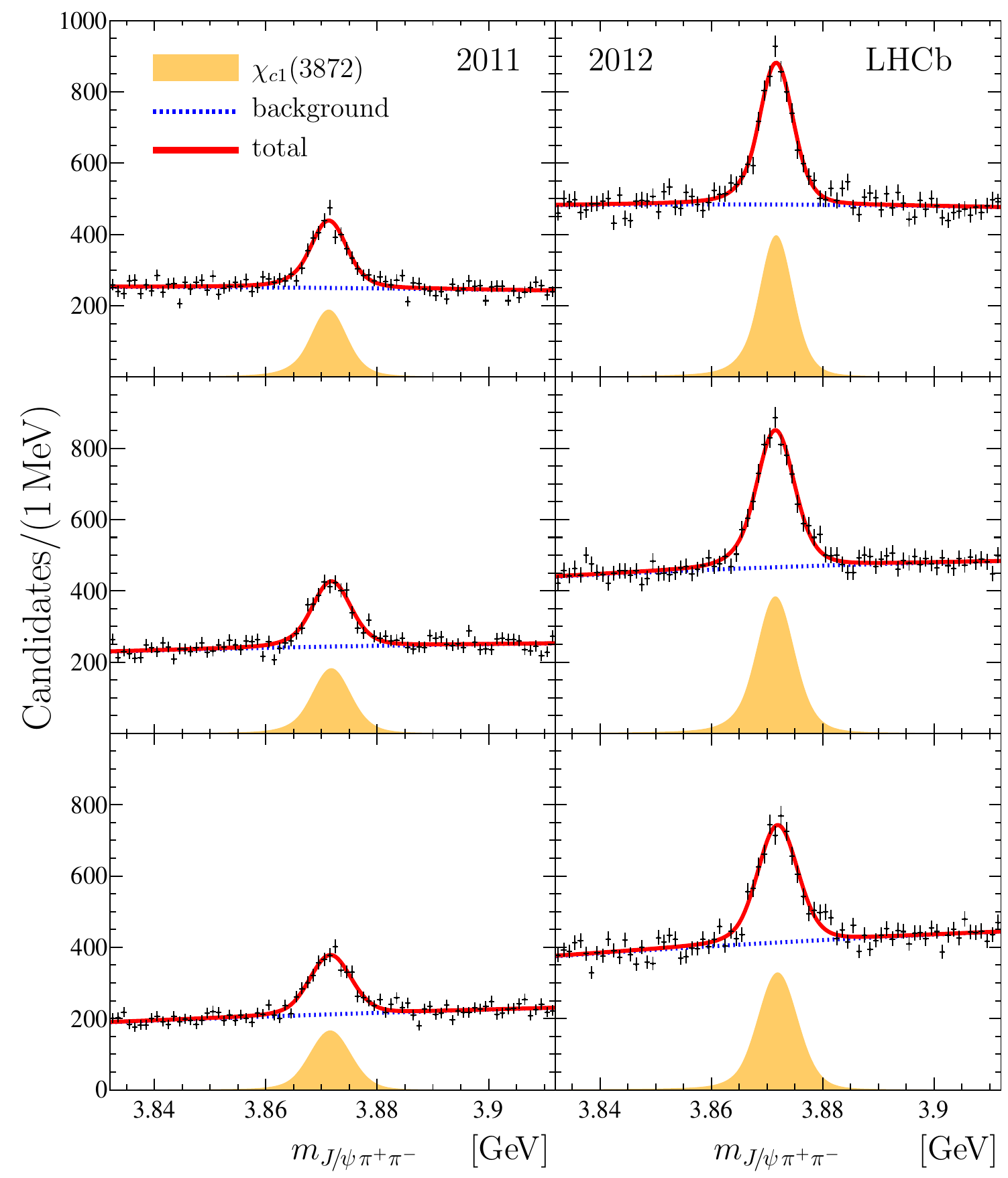}
\caption{\small
  Mass distributions for $\jpsi\pip\pim$~candidates
  in the $\chicone(3872)$ region for (top)\,the low,
  (middle)\,mid and (bottom)\,high $p_{\pip\pim}$~bins.
  The left\,(right)-hand plot is for 2011\,(2012) data.
  The
  projection of the fit described in the text is
  superimposed.
  }
\label{fig:xcbgpolx}
\end{figure*}

 For the six $p_{\pip\pim}$ data samples   the $\jpsi \pip \pim$ mass
 distributions in the $\chicone(3872)$ region and fits are shown
 in Fig.~\ref{fig:xcbgpolx}
 and the results summarised in
Table~\ref{tab:cbgpolxresults}.
Binning the data and calculating the \chisq~probability
of consistency with the fit model
gives values much larger
than $5 \%$ for all bins apart from
the high-momentum bin in the 2012 data where
the probability is $2 \%$. The values of
$\Delta m$ and $\Gamma_{\mathrm{BW}}$ are
consistent between the bins giving confidence in
the results.

\begin{table*}[tb]
  \centering
  \caption{\small Results for $\Delta m$ and $\Gamma_{\mathrm{BW}}$ and
  $\chicone(3872)$~signal yields. The quoted uncertainties are statistical.
    } \label{tab:cbgpolxresults}
  \vspace*{3mm}
\begin{tabular*}{0.80\textwidth}{@{\hspace{2mm}}l@{\extracolsep{\fill}}rccc@{\hspace{2mm}}}
	Year  & $p_{\pip\pim}~\left[\!\gev\right]$\ \
	& $\Delta m~\left[\!\mev\right]$
	& $\Gamma_{\mathrm{BW}}~\left[\!\mev\right]$
	& $N_{\mathrm{sig}}~\left[10^3\right]$
    \\[1.5mm]
    \hline
    \\[-3mm]
    2011~ &  $\phantom{00\le}p_{\pip\pim}<12$
     & $185.32\phantom{0} \pm 0.20\phantom{0}$
     & $1.88 \pm 0.74$
     &  $\phantom{0}1.78\pm0.13$ 
    \\
     2011~ &  $12\le p_{\pip\pim}<20$
      & $185.78\phantom{0} \pm 0.21\phantom{0}$
      & $1.53 \pm 0.74$
      & $\phantom{0}1.79\pm0.13$ 
    \\
    2011~ & $20\le p_{\pip\pim}<50$
     & $185.46\phantom{0} \pm 0.21\phantom{0}$
     & $1.03 \pm 0.82$
     & $\phantom{0}1.68\pm0.13$ 
     \\
     2012~ & $\phantom{00\le}p_{\pip\pim}<12$
     &  $185.63\phantom{0} \pm 0.13\phantom{0}$
     & $1.23 \pm 0.47$
     & $\phantom{0}3.24\pm0.18$ 
    \\
    2012~ & $12\le p_{\pip\pim}<20$
     & $185.47\phantom{0} \pm 0.14\phantom{0}$
     & $1.48 \pm 0.48$
     & $\phantom{0}3.70\pm0.18$ 
    \\
     2012~ & $20\le p_{\pip\pim}<50$
     &  $185.81\phantom{0} \pm 0.15\phantom{0}$
     & $1.15 \pm 0.57$
     & $\phantom{0}3.26\pm0.17$ 
    \\[1.5mm]
    \hline
    \\[-3mm]
    \multicolumn{2}{l}{Total}
    &   $185.588\pm 0.067$
    &   $1.39\pm0.24$
    &  $15.63\pm 0.38$
\end{tabular*}
\end{table*}

A simultaneous fit is made to the six data samples
with  $\Delta m$ and $\Gamma_{\mathrm{BW}}$ as shared
parameters.
This gives $\Delta m = 185.588 \pm 0.067 \mev$
and $\Gamma_{\mathrm{BW}} = 1.39 \pm 0.24
\mev$,
where the uncertainties are statistical. Consistent values are
found when these parameters are determined through a weighted average
of the six individual bins, or by summing the likelihood profiles returned by the fit.

The dominant systematic uncertainty on the
mass difference $\Delta m $
arises from the $3\times 10^{-4}$ relative
uncertainty on the momentum scale.
Its effect is evaluated by
adjusting the four-vectors of the pions by
this amount and repeating the analysis.
The bias on $\Delta m$ from QED
radiative corrections is determined
to be $(-10 \pm 14)\kev$ using the simulation,
which uses \photos~\cite{Golonka:2005pn}  to
model this effect.
The measured value of $\Delta m$
is corrected by this value
and the uncertainty considered as a systematic error. The small
uncertainty on the fitted values of the $\psitwos$ mass
is also propagated to the $\Delta m$~value.
Biases arising from the modelling of the resolution
 and the treatment of the background shape are evaluated to
be $2\kev$ using the discrete profiling
method described in Ref.~\cite{Dauncey:2014xga}. The uncertainties on
the $\Delta m$ measurement are summarised in
Table~\ref{tab:masssyst}. Combining all uncertainties,
the mass
splitting between the $\chicone(3872)$ and $\psitwos$ mesons is determined as
\begin{equation*}
\Delta m = 185.598 \pm 0.067
\pm 0.068\mev\,,
\end{equation*}
where the first uncertainty is statistical and the second is systematic.
The value of $\Delta m$ can be translated into an absolute measurement of the
$\chicone(3872)$ mass using
\mbox{$m_{\psitwos} = 3686.097 \pm 0.010 \mev$} from
Ref. \cite{PDG2019}, yielding
\begin{equation*}
m_{\chicone(3872)} = 3871.695 \pm 0.067
\pm 0.068 \pm 0.010\mev\,,
\end{equation*}
where the third uncertainty is due to
the knowledge of the~\psitwos mass.
For these measurements it is assumed that interference effects with other
partially reconstructed $\bquark$-hadron decays do not affect the lineshape. This
assumption is reasonable since many exclusive $\bquark$-hadron
decays contribute to the final sample, and the $\chicone(3872)$ state is narrow.
This assumption has been explored
in pseudoexperiments varying the composition and phases of
the possible decay amplitudes that are likely to contribute
to the observed data set. These studies conservatively
limit the size of any possible effect on $m_{\chicone(3872)}$ to be less than $40\kev$.
\begin{table}[t]
\centering
\caption{\small Systematic uncertainties on the measurement of the mass difference~$\Delta m$. }
\label{tab:masssyst}
  \vspace{3mm}
  	\begin{tabular*}{\linewidth}{@{\hspace{3mm}}l@{\extracolsep{\fill}}c@{\hspace{3mm}}}
   Source & Uncertainty~$\left[\!\mev\right]$
    \\[1.5mm]
    \hline
    \\[-3mm]
     Momentum scale &  0.066 \\
     Radiative corrections   & 0.014 \\
     Fitted $\psitwos$ mass uncertainty & 0.007 \\
     Signal + background model & 0.002
        \\[1.5mm]
    \hline
    \\[-3mm]
     Sum in quadrature & 0.068
    \end{tabular*}
\end{table}

The uncertainties from the knowledge of $s_f$ and $n$ are already
included in the statistical uncertainty of  $\Gamma_{\mathrm{BW}}$ via the Gaussian
constraints. Their contribution to the statistical uncertainty is estimated to be $0.05 \mev$ by comparison to a fit 
with these parameters fixed. Further uncertainties arise from the
choice of signal and background model. These are evaluated using the
discrete profiling method with the alternative
models described above. 
Based upon these studies
an uncertainty of $0.10\mev$
is assigned. The uncertainty due to possible differences
in the $\pt$ distribution between data
and simulation is evaluated by weighting the simulation
to achieve better agreement and lead to a $0.01
\mev$ uncertainty. Summing these values in quadrature gives a total
uncertainty of $0.1 \mev$.

The value of $\Gamma_{\mathrm{BW}}$, including
systematic uncertainties,
\begin{equation*}
    \Gamma_{\mathrm{BW}} =  1.39 \pm 0.24 \pm 0.10 \mev\,,
\end{equation*}
differs from zero by more than $5$~standard deviations.
Fits were also made fixing
$\Gamma_{\mathrm{BW}}$ to zero and allowing $s_f$ to
float in each bin
without constraint. The value of $s_f$ obtained is between 1.2 and
1.25 depending on the bin, much larger than can be reasonably
explained by differences in the mass resolution between data and
simulation after the calibration using the $\psitwos$ data. 

Care is needed in the interpretation of the measured
$\Gamma_{\mathrm{BW}}$ and $m_{\chicone(3872)}$~parameters  since
\mbox{$\left|m_{\Dz} + m_{\Dstarzb} - m_{\chicone(3872)}\right|
< \Gamma_{\mathrm{BW}}$}.
The Breit--Wigner parameterisation may not
be valid since it neglects the opening of the $\Dz\Dstarzb$~channel.


\section{Flatt\'e model}
\label{sec:hanhart}
\subsection{The Flatt\'e lineshape model}
The proximity of the $\chicone(3872)$ mass to the \Dz\Dstarzb threshold
distorts the lineshape from the simple Breit--Wigner form. This  has to be
taken into account explicitly.
The general solution to this problem requires
a full understanding of the analytic
structure of the coupled-channel scattering amplitude.
However, if the relevant
threshold is close to the resonance,
simplified parametrisations are available
and have been used  to describe the $\chicone(3872)$
lineshape~\cite{PhysRevD.76.034007,Kalashnikova:2009gt}.

In the $\jpsi \pip\pim$ channel the $\chicone(3872)$ lineshape
as a
function of the energy with respect to
the $\Dz\Dstarzb$ threshold, 
$E\equiv m_{\jpsi\pip\pim}- \left(m_{\Dz}+m_{\Dstarz}\right)$,
can  be written as
\begin{equation}
\label{eqn:flatte}
\dfrac{d R(\jpsi\pip\pim)}
{dE}\propto
\dfrac{\Gamma_\Prho(E)}{\left|D(E)\right|^2},
\end{equation}
where
$\Gamma_{\Prho}(E)$ is the contribution of the $\jpsi\pip\pim$
channel to
the width of the $\chicone(3872)$ state.
The complex-valued denominator function, taking into account
the $\Dz\Dstarzb$
and $\Dp\Dstarm$ two-body thresholds,
and the $\jpsi\pip\pim$  $\jpsi\pip\pim\piz$ channels, is given by
\begin{equation}
\label{eqn:denominator}
D(E)=E-E_f+\frac{i}{2}\left[g\left(k_1+k_2\right)+
\Gamma_{\Prho}(E)+\Gamma_{\Pomega}(E)+\Gamma_0\right].
\end{equation}
The Flatt\'e energy parameter, $E_f$,
is related to a mass parameter, $m_0$,
via the relation \mbox{$E_f=m_0-(m_{\Dz}+m_{\Dstarz})$}.
The width $\Gamma_0$ is introduced
in Ref.~\cite{Kalashnikova:2009gt} to represent further
open channels, such as radiative decays.
The model assumes an isoscalar
assignment of the $\chicone(3872)$ state,
using the same effective coupling, $g$,
for both channels. The relative momenta of the decay products in
the rest frame of the two-body system, $k_1$ for  $\Dz\Dstarzb$
and $k_2$ for the
$\Dp\Dstarm$ channel, are given by
  \begin{equation}
  \label{eqn:breakup}
  k_1=\sqrt{2\mu_1E},\ \quad
   k_2=\sqrt{2\mu_2(E-\delta)} \,,
  \end{equation}
where $\delta = 8.2\mev$ is the isospin splitting
between the two channels. The reduced masses are given by
\mbox{$\mu_1=\tfrac{m_{\Dz}m_{\Dstarz}}{ (m_{\Dz} + m_{\Dstarz})}$} and
\mbox{$\mu_2=\tfrac{m_{\Dp}m_{\Dstarm}}{ (m_{\Dp} + m_{\Dstarm})}$}.
For $m_{\jpsi\pip\pim}$ masses below the $\Dz\Dstarzb$ and $\Dp\Dstarm$
thresholds these momenta become
imaginary and thus their contribution to the denominator
will be real.
The  energy dependence  of the $\jpsi\pip\pim$ and $\jpsi\pip\pim\piz$ partial widths
is given by~\cite{Kalashnikova:2009gt}
\begin{eqnarray}
\Gamma_{\rho}(E)= & f_\Prho \int\limits_{2m_\Ppi}^{M(E)}\dfrac{dm'}{2\pi}
\dfrac{q(m',E)\,\Gamma_\rho}{(m'-m_\rho)^2+\Gamma_\rho^2/4} \,, \label{eqn:twobodyint1} \\
\Gamma_{\omega}(E)= &f_\omega \int\limits_{3m_\pi}^{M(E)}
\dfrac{dm'}{2\pi}
\dfrac{q(m',E)\,\Gamma_\omega}{(m'-m_\omega)^2+\Gamma_\omega^2/4} \,. \label{eqn:twobodyint2}
\end{eqnarray}
The known values for masses $m_{\rho}$,
$m_{\omega}$ and widths $\Gamma_{\rho}$, $\Gamma_{\omega}$~\cite{PDG2019} are
used and the lineshapes are approximated
with fixed-width Breit--Wigner functions.
The partial widths are parameterised  by the respective
effective couplings $f_\Prho$ and $f_{\Pomega}$
and the phase space of these decays,
where intermediate resonances $\decay{\Prho^0}{\pip\pim}$
and $\decay{\Pomega}{\pip\pim\piz}$ are
assumed.
The dependence on $E$ is given by the upper boundary
of the integrals \mbox{$M(E)=E+(m_{\Dz}+m_{\Dstarz})-m_{\jpsi}$}.
The momentum of the two- or three-pion system
in the rest frame of the $\chicone(3872)$ is given by

\begin{equation}
  q(m',E)=
  \sqrt{\dfrac{\left[M^2(E)-(m'+m_{\jpsi})^2\right]\left[M^2(E)-(m'-m_{\jpsi})^2\right]}{4M^2(E)}}.
\end{equation}

The model as specified contains five free parameters: $m_0,g,\Gamma_0$
and the effective couplings $f_\Prho$ and $f_{\Pomega}$.
In contrast to the Breit--Wigner lineshape, the parameters
of the Flatt\'e model cannot be easily interpreted in terms
of the mass and width of the state. Instead it is necessary
to determine the location of the poles of the amplitude.
The analysis proceeds with a fit of the Flatt\'e amplitude
 to the data 
 and subsequent search for the poles. 

The resulting Flatt\'e lineshape replaces the
Breit--Wigner function
and is convolved with the resolution
models described in the first part of the paper. 
The Flatt\'e parameters are estimated from a simultaneous
unbinned likelihood fit to the $\jpsi \pip \pim$ mass distribution
in the six $p_{\pip\pim}$ data samples. 
The data points
are corrected for the observed shifts of the reconstructed mass
of the \psitwos in each bin.

\subsection{Fits of the Flatt\'e lineshape to the data}\label{sec:flattefit}

In order to obtain stable results when using the
coupled channel model to describe
the $\jpsi\pip\pim$ mass spectrum, a relation between
the effective couplings
$f_\rho$ and $f_{\Pomega}$ is imposed.
This relation
requires that the branching fractions
of the $\chicone(3872)$~state
to $\jpsi\Prho^0$ and $\jpsi\Pomega^0$~final states
are equal,
 which is consistent
 with experimental
 data~\cite{Aubert:2008g,Adachi:2008sua,Choi:2011fc},
 thus eliminating one free parameter
 in the fit.
  Furthermore, a Gaussian
  fit constraint is applied on the ratio of branching fractions
 \begin{equation}
R_{\D\Dstarb}=\dfrac{\decay{\Gamma(\chicone(3872)}{\jpsi\pip\pim})}
{\Gamma( \decay{\chicone(3872)}{\Dz\Dstarzb})}=0.11\pm0.03\,. \label{eq:rdd}
 \end{equation}
 The value used here is obtained as the weighted
average of the results from the
BaBar~\cite{Aubert:2008g} and
Belle~\cite{Adachi:2008sua,Choi:2011fc} collaborations,
as listed in Ref.~\cite{CHEN20161}.
The Flatt\'e model reduces to the Breit--Wigner model
as a special case, namely
when there is no additional decay channel
available near the resonance. However,
the $R_{\D\Dstarb}$ constraint enforces
a large coupling to the $\Dz\Dstarzb$
channel and the lineshape will
be different from the Breit--Wigner function in the
region of interest.

For large couplings to the two-body channel the Flatt\'e parameterisation
exhibits a scaling property~\cite{Baru2005} that
prohibits the unique determination of all
free parameters on the given data set.
Almost identical lineshapes are obtained when the
parameters $E_f$, $g$, $f_\rho$ and $\Gamma_0$ are scaled appropriately.
In particular, it is possible to
counterbalance a lower value of $E_f$ with a
linear increase in
the coupling to the $\D\Dbar^*$~channels $g$. While this is not a true
symmetry of the parameterisation --- there are subtle differences in
the tails of the
lineshape --- in practice, within the experimental precision  this
effect leads to
strong correlations between the parameters.

\begin{figure}[t]
  \centering
  \includegraphics[width=1.0\columnwidth]{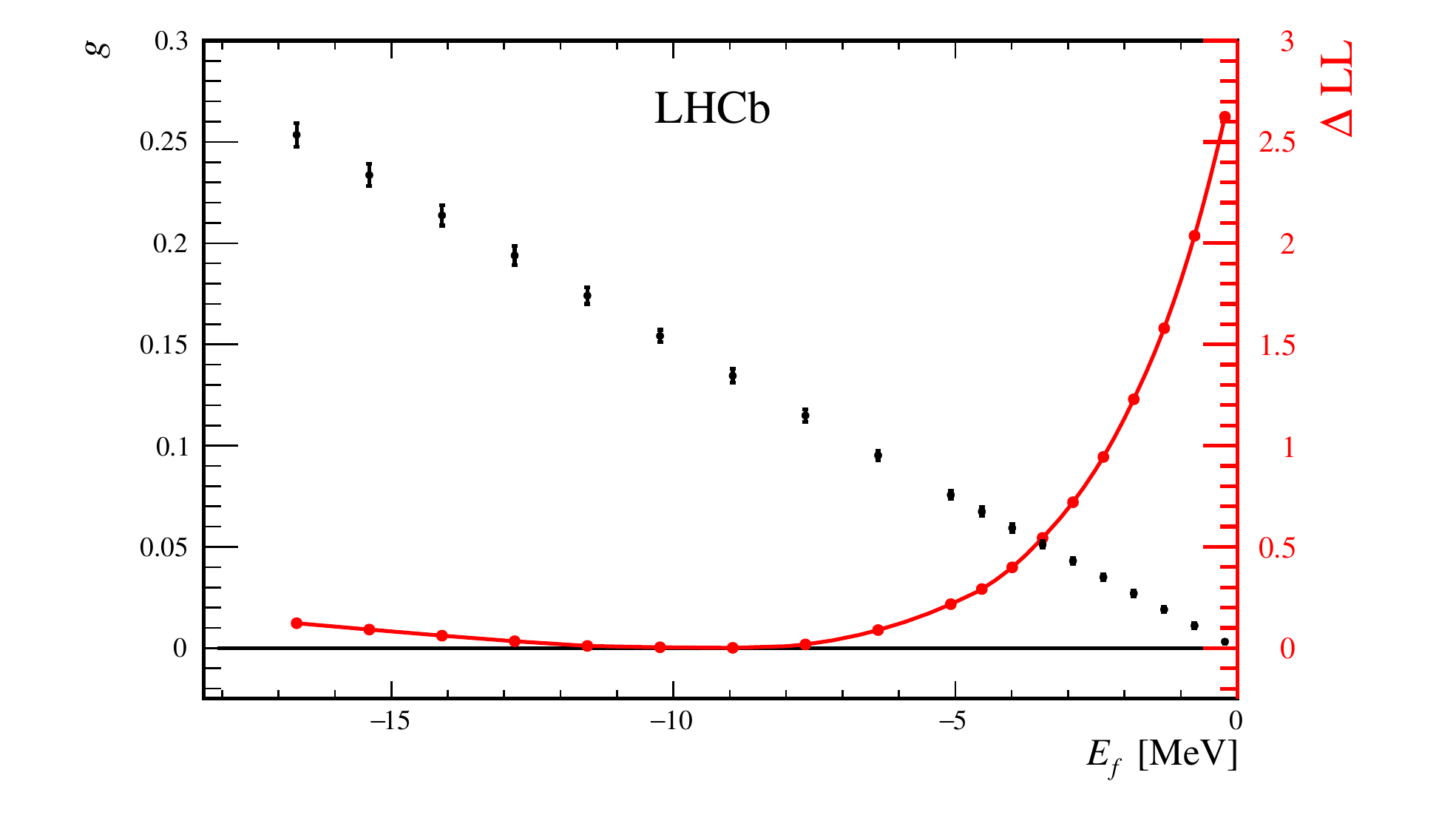}
  \caption{\small The coupling to the
    $\D\Dbar^*$ channels $g$ as a function  of
    Flatt\'e energy parameter $E_f$ (black points with error bars).
    The corresponding change in negative log likelihood, $\Delta\mathrm{LL}$
    is shown as well (red dots).}
  \label{fig:scanLL}
\end{figure}

Figure~\ref{fig:scanLL} illustrates the scaling behavior in the data.
The black points show the best-fit result for the parameter
$g$ evaluated at fixed $E_f$, optimising the remaining parameters
at every step.
To a good approximation $g$ depends linearly
on $E_f$ with
\begin{equation}
\dfrac{d g}{ dE_f}=
(-15.11\pm0.16)\gev^{-1}\,.
\end{equation}
The red points show the negative log likelihood
relative to is minimum value
 $\Delta\mathrm{LL}$ for each of
these fits, revealing a shallow minimum around $m_0=3860\mev$.
At lower $E_f$ values  $\Delta\mathrm{LL}$  raises very slowly,
reaching a value of 1 around $-270\mev$. Values of $E_f$ approaching
the $\Dz\Dstarzb$
threshold are disfavoured, though. In particular good quality
fits are obtained only for negative values of $E_f$.
A similar phenomenon has been observed in the previous
analyses of \babar  and \belle data and is discussed
in Ref.~\cite{Kalashnikova:2009gt}.
As in those studies, for the remainder of the paper
the practical solution of fixing \mbox{$m_0=3864.5\mev$},
corresponding to $E_f=-7.2\mev$, is adopted. The
remaining model parameters are evaluated with
this constraint applied. This  procedure has been
validated using pseudoexperiments
and no significant bias is found.
For $g$  and $\Gamma_0$ the uncertainties
estimated by the fit agree with the spread
of the pseudoexperiments.
For $f_\rho$  an
uncertainty which is 10\% larger than what is found in the
pseudoexperiments is observed and this
conservative estimate is reported.
The measured values for $g$, $f_\rho$  and $\Gamma_0$
are presented in Table~\ref{tab:finalflatte}.
In order to fulfill the constraint on the branching
ratios, Eq.~\eqref{eq:rdd},
the effective coupling, $f_\omega$, is found
to be $0.01$.

\begin{table}[tb]
\centering
\caption{\small Results from the constrained Flatt\'e fit.
The uncertainties are statistical.
}
\label{tab:finalflatte}
\vspace{2mm}
\begin{tabular*}{\columnwidth}{@{\hspace{3mm}}c@{\extracolsep{\fill}}ccc@{\hspace{3mm}}}
   $g$
   & $f_\rho \times 10^3$
   & $\Gamma_0\,\left[\!\mev\right]$
   & $m_0\,\left[\!\mev\right]$
    \\[1.5mm]
    \hline
    \\[-3mm]
     $0.108\pm0.003$ 
  &  $1.8\pm0.6$ 
  & $1.4\pm0.4$   
  & $3864.5\,{\mathrm{(fixed)}}$
\end{tabular*}
\end{table}

\begin{table*}[tb]
\centering
\caption{\small Systematic uncertainty on the measurement of the Flatt\'e parameters.
}
\label{tab:flattesystematics}
\vspace{2mm}
   \begin{tabular*}{0.7\textwidth}{@{\hspace{2mm}}l@{\extracolsep{\fill}}cccccc@{\hspace{2mm}}}
    Systematic
    & \multicolumn{2}{c}{$g$}
    & \multicolumn{2}{c}{$f_\Prho \times 10^3$}
    & \multicolumn{2}{c}{$\Gamma_0~\left[\!\mev\right]$}
    \\[1.5mm]
    \hline
    \\[-3mm]
  Model
  & $~~+0.003$          & $-0.004~~$
  & $~~+0.6 $          & $-0.5~~$
  & $~~+0.5$ & $-0.4~~$
  \\ 
  Momentum scale
  & $~~+0.003$          & $-0.003~~$
  & $~~+0.1$ & $-0.2~~$
  & $~~+0.1$ & $-0.2~~$
  \\ 
  Threshold mass
  & $~~+0.003 $          & $-0.003~~$
  & $~~+0.2 $ & $-0.2~~$
  & $~~+0.2 $ & $-0.3~~$
  \\ 
  \Dstarz width
  &   & $-0.001~~$
  &   &
  &   & $-0.2~~$
   \\[1.3mm]
    \hline
    \\[-3mm]
  Sum in quadrature
  & $~~+0.005$ & $-0.006~~$
  & $~~+0.7$  & $-0.6~~$
  & $~~+0.6$  & $-0.6~~$
  \end{tabular*}
\end{table*}

The systematic uncertainties
on the Flatt\'e parameters are summarised
in Table~\ref{tab:flattesystematics} and discussed below.
The systematic uncertainties introduced by the background and
resolution parameterisations
are evaluated in the same way as for the Breit--Wigner analysis,
using discrete profiling.
The impact of the momentum scale uncertainty is investigated by
shifting the data points
by $66\kev$ and repeating the fit.
Further systematic uncertainties are particular
to the Flatt\'e parameterisation.
The location of the $\Dz\Dstarzb$~threshold  is known to a
precision of $0.11\mev$~\cite{PDG2019}.
Varying the threshold  by this amount and repeating the fit
leads to an uncertainty on the parameters which is similar to
that introduced by the momentum scale.
Finally, the $\Dstarz$ meson has a finite natural width, for which an
upper limit of $\Gamma_{\Dstarz}<2.1\mev$~\cite{PDG2019} has been measured.
However, theoretical predictions estimate $\Gamma_{\PD^*}=65.5\pm15.4\,\kev$ \cite{Braaten:2007dw},
based on the measured width of the $\Dstarp$~meson.
Modified lineshape models taking into account
the finite width of the $\Dstarz$
are available. In particular,
Refs.~\cite{Braaten:2007dw, PhysRevD.81.094028}
suggest  replacing
$k_1(E)$ in Eq.~\eqref{eqn:breakup} with
\begin{equation}
k^{\prime}_1(E) = \sqrt{-2\mu\left(E-E_R + i\Gamma_{\Dstarz}/2\right)}\,,
\end{equation}
where
\mbox{$E_R \equiv m_{\Dstarz} - m_{\Dz} - m_{\piz}$}.
The reduced mass, $\mu$, is
calculated as
\mbox{$\tfrac{m_{\Dz}(m_{\Dz} + m_{\piz})}{  (2m_{\Dz} + m_{\piz})}$}.
With this modification there is always a contribution to both the
imaginary and
real part of the denominator function in Eq.~\eqref{eqn:denominator}.
Repeating the fit results in a similar but worse
fit quality with a log-likelihood difference of $0.1$.
The width $\Gamma_0$ is reduced by $0.2\,\mev$,
which is the smallest systematic uncertainty on this parameter.

\subsection{Comparison between Breit--Wigner
and Flatt\'e lineshapes}

\begin{figure}[t]
\includegraphics*[width=\columnwidth]{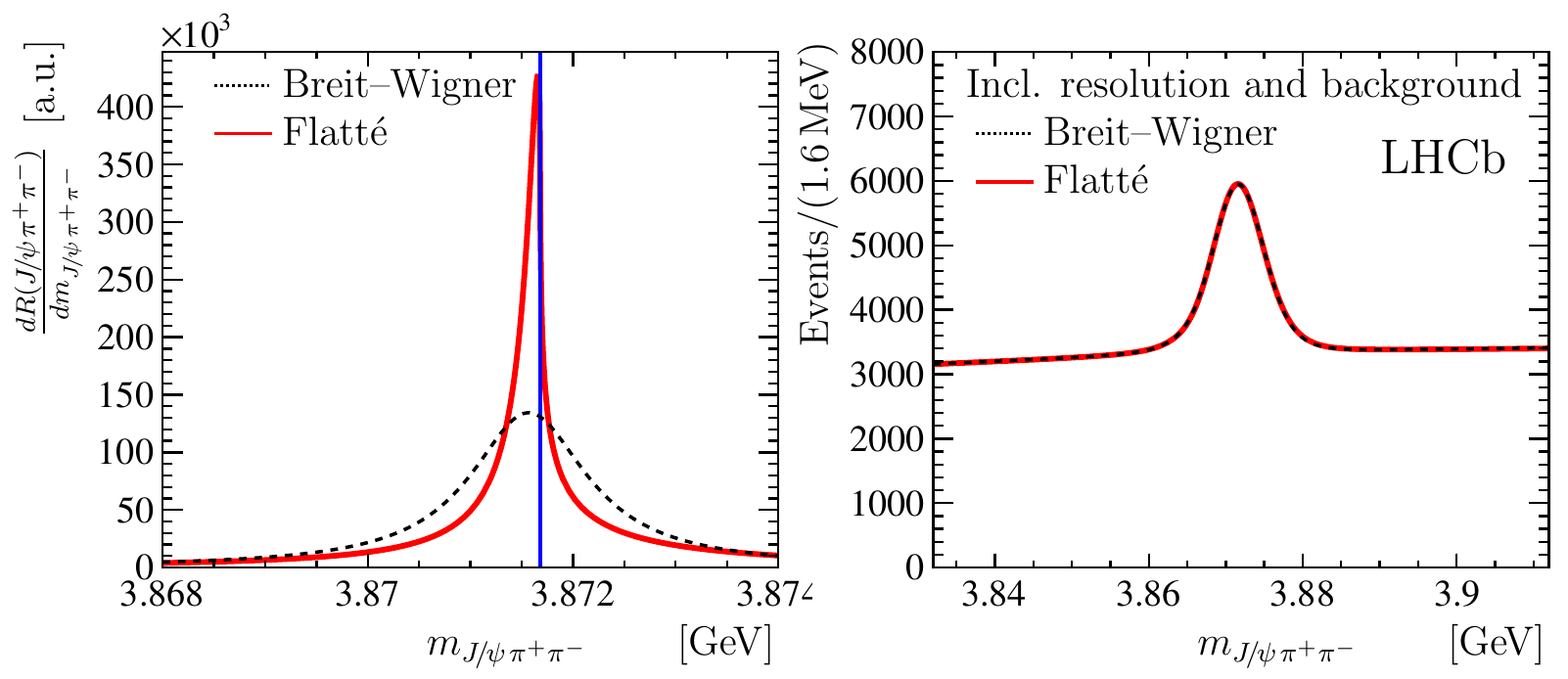}
\caption{\small
  Comparison of the Flatt\'e\,(solid, red)
  and Breit--Wigner\,(dotted, black) lineshapes.
  The left plot shows the raw lineshapes for
  the default fits.
  The location of the \Dz\Dstarzb threshold is
  indicated by the blue vertical line. On the right
  the distributions are shown after applying smearing
  with the resolution function and adding background.}
\label{fig:linecomp}
\end{figure}

Figure~\ref{fig:linecomp} shows the
comparison between the Breit--Wigner
and the Flatt\'e lineshapes. While in both cases
the signal peaks at the same mass,
the Flatt\'e model results in a signifcantly narrower
lineshape. However, after folding with the
resolution function and adding the
background,
the observable distributions are indistinguishable.

\begin{table}[tb]
\centering
\caption{\small Results of the fit with the
 Flatt\'e lineshape
including statistical and systematic uncertainties.
The Flatt\'e mass parameter $m_0=3864.5\mev$ is used.
}
\label{tab:finalshapeflatte}
\vspace{3mm}
\begin{tabular*}{\columnwidth}{@{\hspace{2mm}}c@{\extracolsep{\fill}}cc@{\hspace{2mm}}}
     Mode~$\left[\!\mev\right]$
  & Mean~$\left[\!\mev\right]$
  & FWHM~$\left[\!\mev\right]$
    \\[1.5mm]
    \hline
    \\[-3mm]
$3871.69^{\,+\,0.00\,+\,0.05}_{\,-\,0.04\,-\,0.13}$ &
~$3871.66^{\,+\,0.07\,+\,0.11}_{\,-\,0.06\,-\,0.13}$ &
$0.22^{\,+\,0.06\,+\,0.25}_{\,-\,0.08\,-\,0.17}$
\end{tabular*}
\end{table}

\begin{figure}[t]
  \includegraphics[width=\columnwidth]{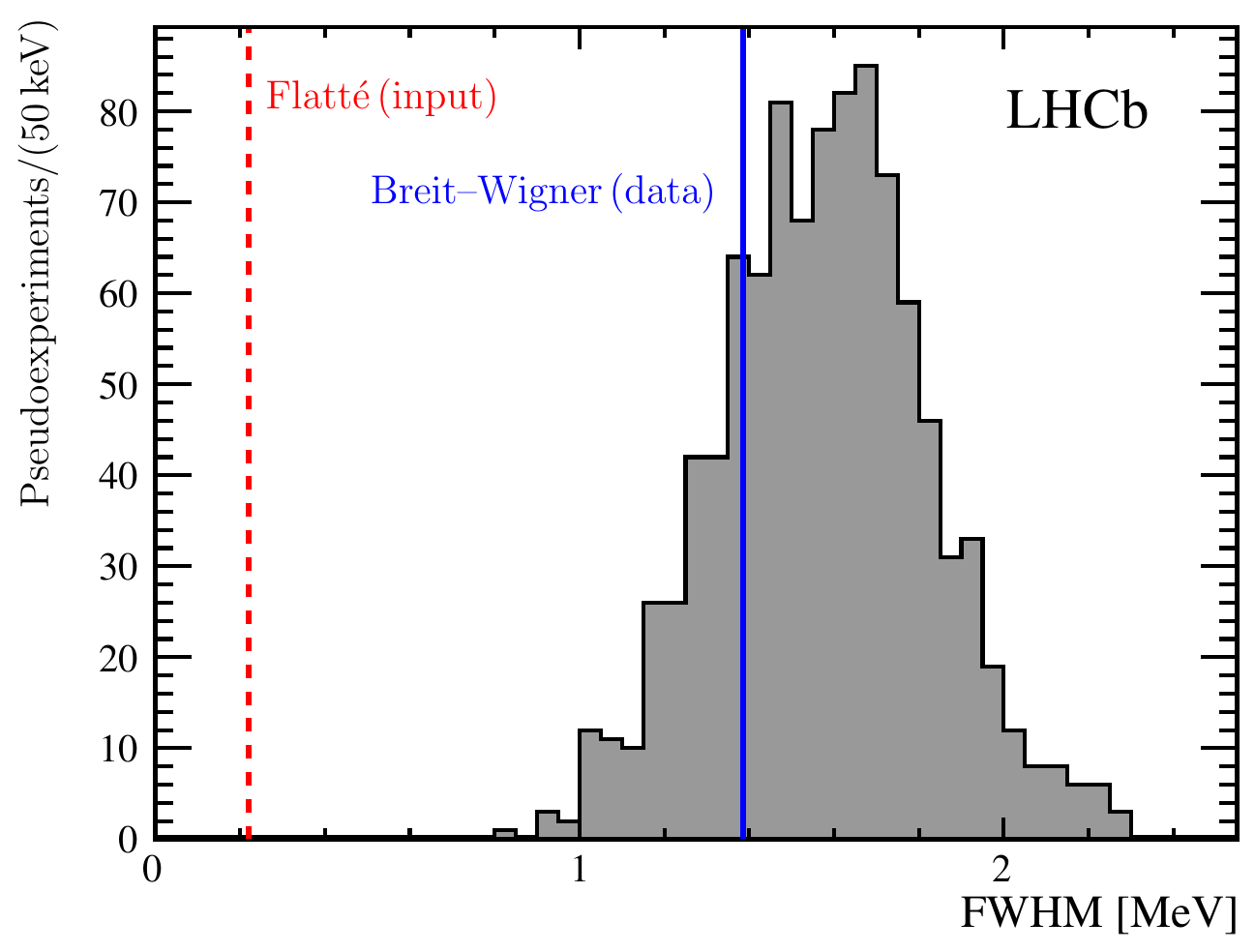}
  \caption{\small
    Distribution of the FWHM obtained for
    simulated experiments
    generated from the result of the Flatt\'e model
    and fitted with the~Breit--Wigner model\,(filled histogram).
    Both models account for the experimental resolution.
    The dashed red line shows the FWHM of the Flatt\'e lineshape,
    while the solid blue line indicates
    the~value of the~Breit--Wigner width observed in data.}
  \label{fig:bwfittoys}
\end{figure}

To quantify this comparison the fit results
for the mode,
the mean and the full width
at half\nobreakdash-maximum\,(FWHM) of the Flatt\'e model and their
uncertainties are summarised in
Table~\ref{tab:finalshapeflatte}. The mode of the
Flatt\'e distribution agrees within
uncertainties with the
Breit--Wigner solution. 
However,
the FWHM of the Flatt\'e model
is a factor of five smaller than the Breit--Wigner width. To check
the consistency of these seemingly
contradictory results, pseudoexperiments
generated with the Flatt\'e model and folded
with the known resolution function
are analysed with the Breit--Wigner model.
Figure~\ref{fig:bwfittoys} shows the resulting distribution
of the Breit--Wigner width determined
from the pseudoexperiments, which is in good agreement
with the value observed in the data.
This demonstrates that the value obtained for
the Breit--Wigner width, after taking into account
the experimental resolution, is consistent with
the expectation of the Flatt\'e model.
The result highlights the importance of
a proper lineshape parameterisation for a measurement
of the location of the pole.

\subsection{Pole search}\label{sec:poles}
The amplitude as a function of the energy defined
by Eq.~\eqref{eqn:denominator}
 can be continued analytically to complex
 values of the energy $E$. This continuation is valid up
to singularities of the amplitude.
There are two types of singularities, which are relevant here:
poles and branch points. Poles of the amplitude in the complex
energy plane are identified with hadronic states. The pole location
is a unique property of the respective state, which is independent
of the production process and the observed decay mode. In the
absence of nearby thresholds the real part of the pole is located
at the mass of the hadron and the imaginary part at half the width
of the state. Branch point singularities occur at the threshold of
every coupled channel and lead to branch cuts in the Riemann surface
on which the amplitude is defined. Each
branch cut corresponds to two Riemann sheets.
Through Eq.~\eqref{eqn:denominator} the amplitude
will inherit the analytic structure of
the square root functions of Eq.~\eqref{eqn:breakup}
that describe the momenta
of the decay products in the rest frame of the two-body system.
The square root is a two-sheeted function of complex energy. In the following,
a convention is used where the two sheets are connected along the negative real axis.
An introduction to this subject
can be found in Refs.~\cite{Eden:1966dnq,Martin:1970xx,Collins:1977xx}
and a  summary is available in Ref.~\cite{PDG2019Resonances}.

For the $\chicone(3872)$~state only the
Riemann sheets associated to the
$\Dz\Dstarzb$~channel are important, since all other thresholds are far from the signal region.
The following convention is adopted to label the relevant sheets:
 \begin{enumerate}
 \item[I:] $E-E_f-\dfrac{g}{2}\left(+\sqrt{-2\mu_1E}+\sqrt{-2\mu_2(E-\delta)}\right)+
 \dfrac{i}{2}\Gamma(E)$
 with $\operatorname{Im} E > 0$\,,
 \item[II:] $E-E_f-\dfrac{g}{2}\left(+\sqrt{-2\mu_1E}+\sqrt{-2\mu_2(E-\delta)}\right)+\dfrac{i}{2}\Gamma(E)$ with $\operatorname{Im} E < 0$\,,
  \item[III:] $E-E_f-\dfrac{g}{2}\left(-\sqrt{-2\mu_1E}+\sqrt{-2\mu_2(E-\delta)}\right)+\dfrac{i}{2}\Gamma(E)$ with $\operatorname{Im} E < 0$\,,
  \item[IV:] $E-E_f-\dfrac{g}{2}\left(-\sqrt{-2\mu_1E}+\sqrt{-2\mu_2(E-\delta)}\right)+\dfrac{i}{2}\Gamma(E)$ with $\operatorname{Im} E > 0$\,,
\end{enumerate}
where $\Gamma(E)\equiv \Gamma_{\Prho}(E)+ \Gamma_{\Pomega}(E)+\Gamma_0$.
The fact that the model contains several coupled
channels in addition to the $\Dz\Dstarzb$ channel
complicates the analytical structure.
The sign in front of the momentum $\sqrt{-2\mu_1E}$ is the same for sheets I
and II  and therefore they belong to a single
sheet with respect to the $\Dz\Dstarzb$ channel.
The two regions are labelled separately
due to the presence of the the $\jpsi\pip\pim$,
$\jpsi\pip\pim\piz$ channels,
as well as radiative decays. Those channels have their
associated branch points at smaller masses than the
signal region. The analysis is performed close to
the $\Dz\Dstarzb$ threshold
and points above and
below the real axis lie on different
sheets with respect to those open channels.

Sheets\,I and~II correspond to a physical sheet with
respect to the $\Dz\Dstarzb$ channel, where the amplitude is evaluated in order to compute
the measurable lineshape at real energies $E$. Sheets\,III and~IV correspond to an
unphysical sheet with respect to that channel. Sheet\,II is analytically
connected to sheet\,IV along the real axis,
above the $\Dz\Dstarzb$ threshold.

In the single-channel case, a bound $\Dz\Dstarzb$ state would
appear below threshold on the real axis and on the physical sheet.

A virtual state
would appear as well below threshold on the real axis,
but on the unphysical sheet.
A resonance would appear on the unphysical sheet
in the complex plane \cite{Eden:1966dnq,Martin:1970xx,Collins:1977xx}.
The presence of inelastic, open channels
shifts the pole into the complex plane and turns both a bound
state as well as a virtual state into resonances.
In the implementation of the amplitude used for
the analysis, the branch cut for the
$\Dz\Dstarzb$~channel is taken to go from threshold towards larger energy $E$,
while the branch cuts associated to the open channels
$\Gamma(E)$ are chosen to lie along the negative real axis.
The analytic structure around the branch cut associated to the~$\Dp\Dstarm$
threshold is also investigated, but no nearby
poles are found on the respective Riemann sheets.

At the best estimate of the Flatt\'e parameters the model exhibits two pole singularities.
The first pole appears on
sheet\,II and is located very close to the
$\Dz\Dstarzb$ threshold .
The location of this pole with respect to the branch point,
obtained using the algorithm described in Ref.~\cite{PoleFinder}, is
$E_{\mathrm{II}}=(0.06-0.13\:i)\mev$.
Recalling that the imaginary part
of the pole position corresponds to half the visible width,
it is clear that this pole is responsible for the peaking region of the lineshape.
A second pole is found on sheet\,III. It appears well
below the threshold and
is also further displaced from the physical axis at
\mbox{$E_{\mathrm{III}}=(-3.58-1.22\:i)\mev$}.

\begin{figure}[t]
\centering
\includegraphics[width=1.0\columnwidth]{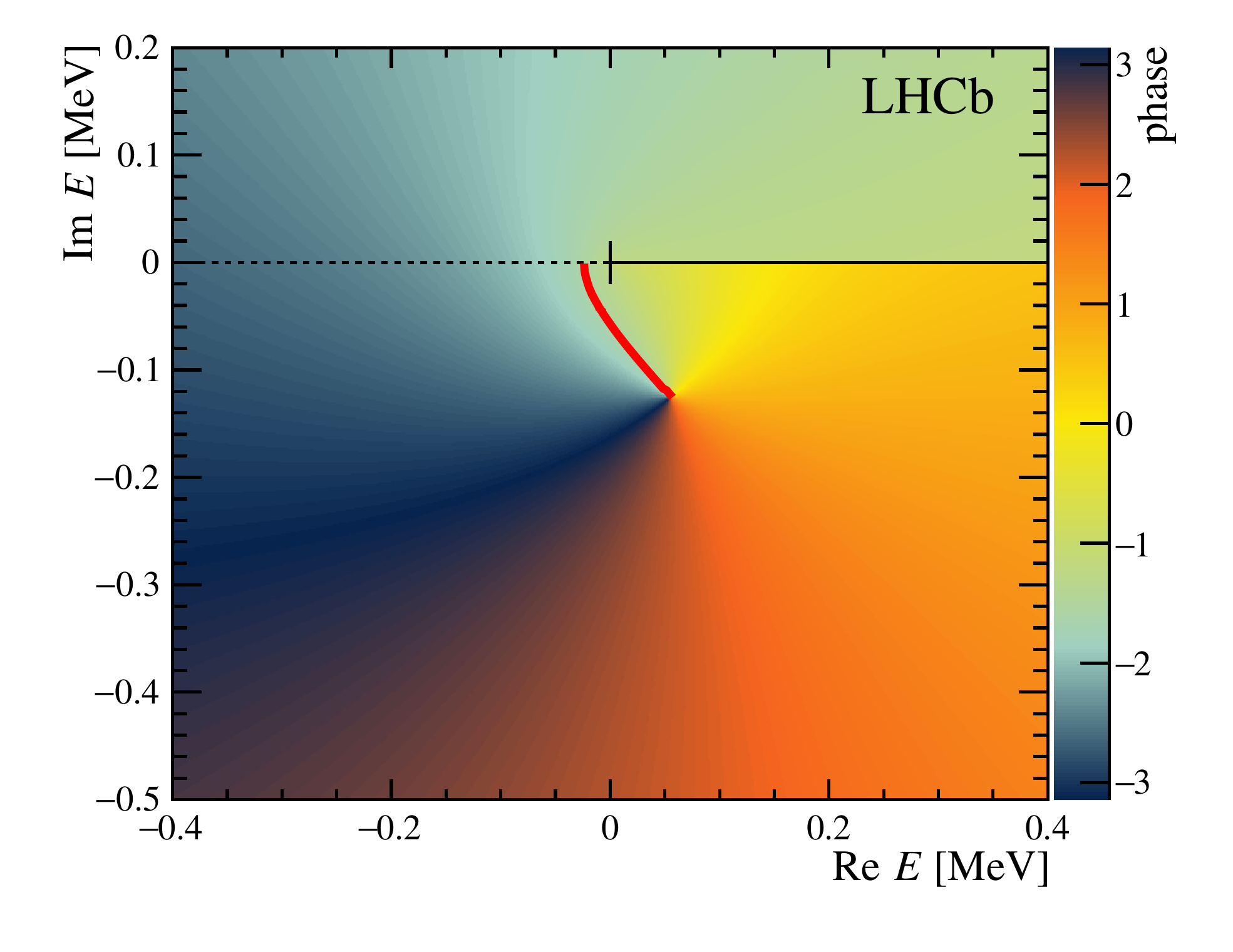}
  \caption{\small The phase of the Flatt\'e amplitude obtained from  the
  fit to the data with \mbox{$m_0=3864.5\mev$}
  on sheets\,I~(for $\operatorname{Im} E>0$)
  and II~(for $\operatorname{Im} E<0$)
  of the complex energy plane.
  The pole singularity is visible at
  $E_{\mathrm{II}}=(0.06  - 0.13\:i)\mev$.
  The branch cut is
  highlighted with the black line.
  The trajectory of the pole taken when
  the couplings to all but the $\D\Dstarb$
  channel are scaled down to zero is indicated in red.
  }
\label{fig:flatteriemannresult}
\end{figure}

Figure~\ref{fig:flatteriemannresult} shows the analytic
structure of the Flatt\'e amplitude
in the vicinity of the threshold. The color code
corresponds to the phase of the amplitude on
sheets\,I~(for $\operatorname{Im} E >0$)
and II~(for $\operatorname{Im} E<0$)
in the complex energy plane. The pole on
sheet\,II is visible, as is the
discontinuity along the  $\Dz\Dstarzb$~branch cut,
which for clarity is also indicated by the black line.
The trajectory followed by the pole when taking
the limit where the couplings to all channels
but $\Dz\Dstarzb$ are sent to zero is shown
in red and discussed below.

\begin{figure}[t]
  \centering
  \includegraphics[width=1.0\columnwidth]{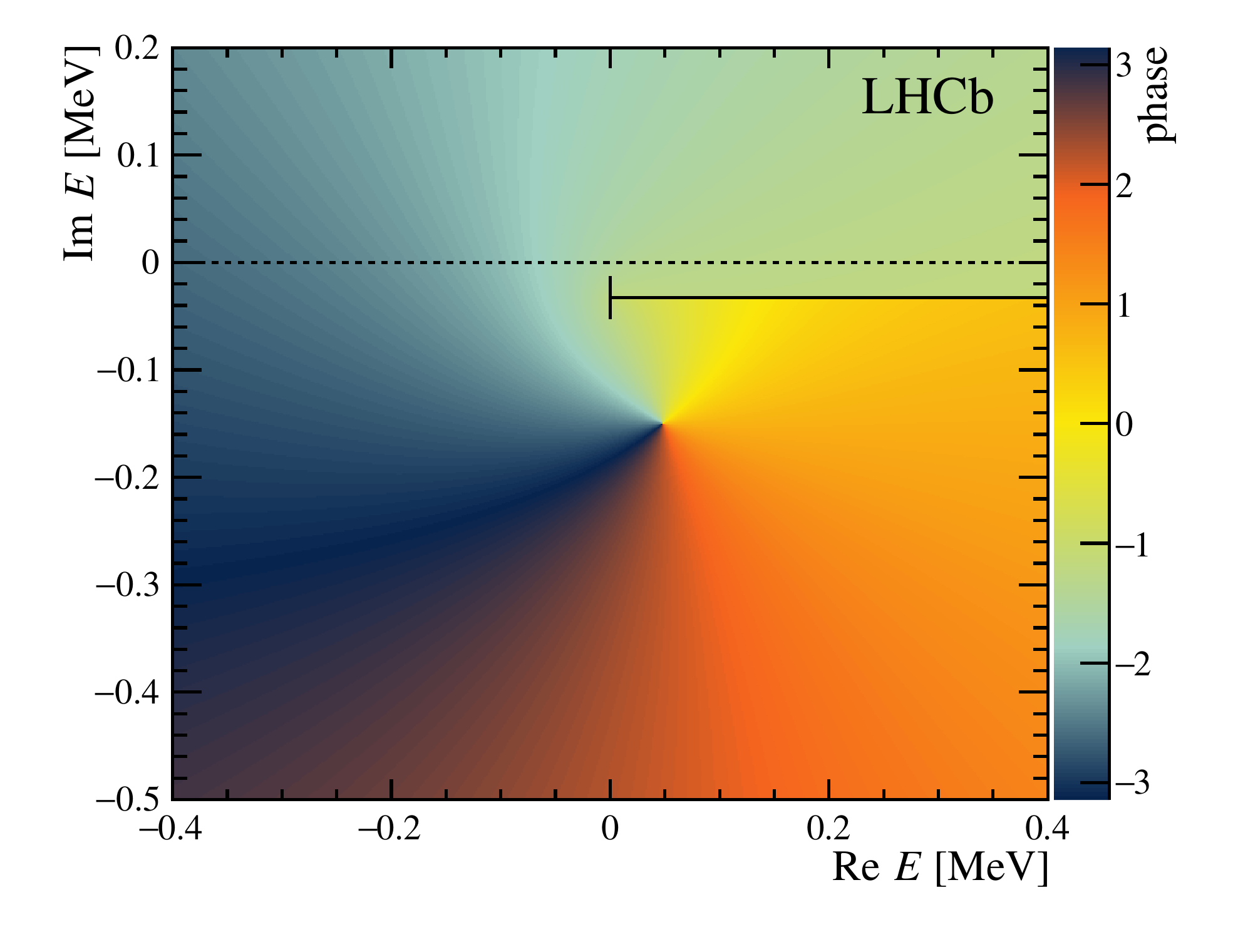}
  \caption{\small The phase of the Flatt\'e amplitude
    as obtained from the fit with a finite \Dstarz width
    of $\Gamma_\Dstarz=65.5\kev$
    on sheets\,I~(for $\operatorname{Im} E>-\Gamma_{\Dstarz}/2$)
    and       II~(for $\operatorname{Im} E<-\Gamma_{\Dstarz}/2$)
    of the complex energy plane.
    Since the \Dstarzb meson is treated as an unstable particle,
    the \Dz\Dstarzb branch cut, indicated by the black solid line, is located at
    $\operatorname{Im} E=-\Gamma_{\Dstarz}/2$.
    The  location of the pole is on the physical
    sheet with respect to the \Dz\Dstarzb system.
  }
\label{fig:polewithDstwidth}
\end{figure}

\begin{figure}[tb]

  \includegraphics*[width=\columnwidth]{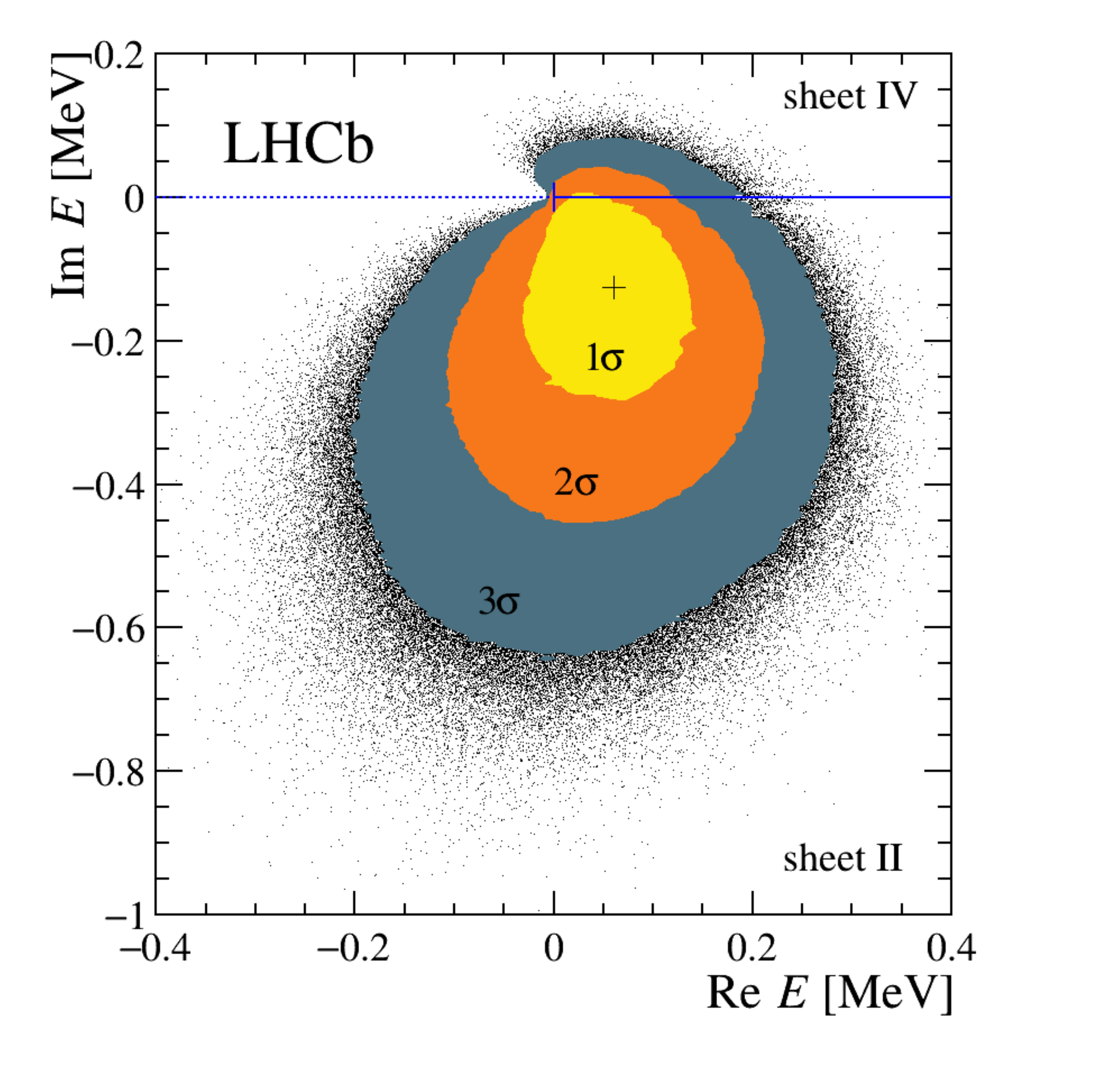}
  \caption{\small Confidence regions for the pole position
   on sheets\,II and IV in the complex energy plane. The displayed uncertainties
   include statistical contributions and the modeling uncertainty.
   The poles are extracted at a Flatt\'e mass point of $m_0=3864.5\mev$.
   The shaded areas are the 1, 2 and 3$\sigma$ confidence regions.
   The branch cut is shown as the blue line.
     The location of the branch cut singularity is
    indicated with a vertical bar at $E=0+0\:i$.
    The best estimates for the pole position is indicated by a cross.
    The black points indicate the samples from
    the pseudoexperiments procedure that lie outside the $3\sigma$ region.
  }
  \label{fig:polesII}
 \end{figure}
 
\begin{figure}[tb]
 \includegraphics[width=\columnwidth]{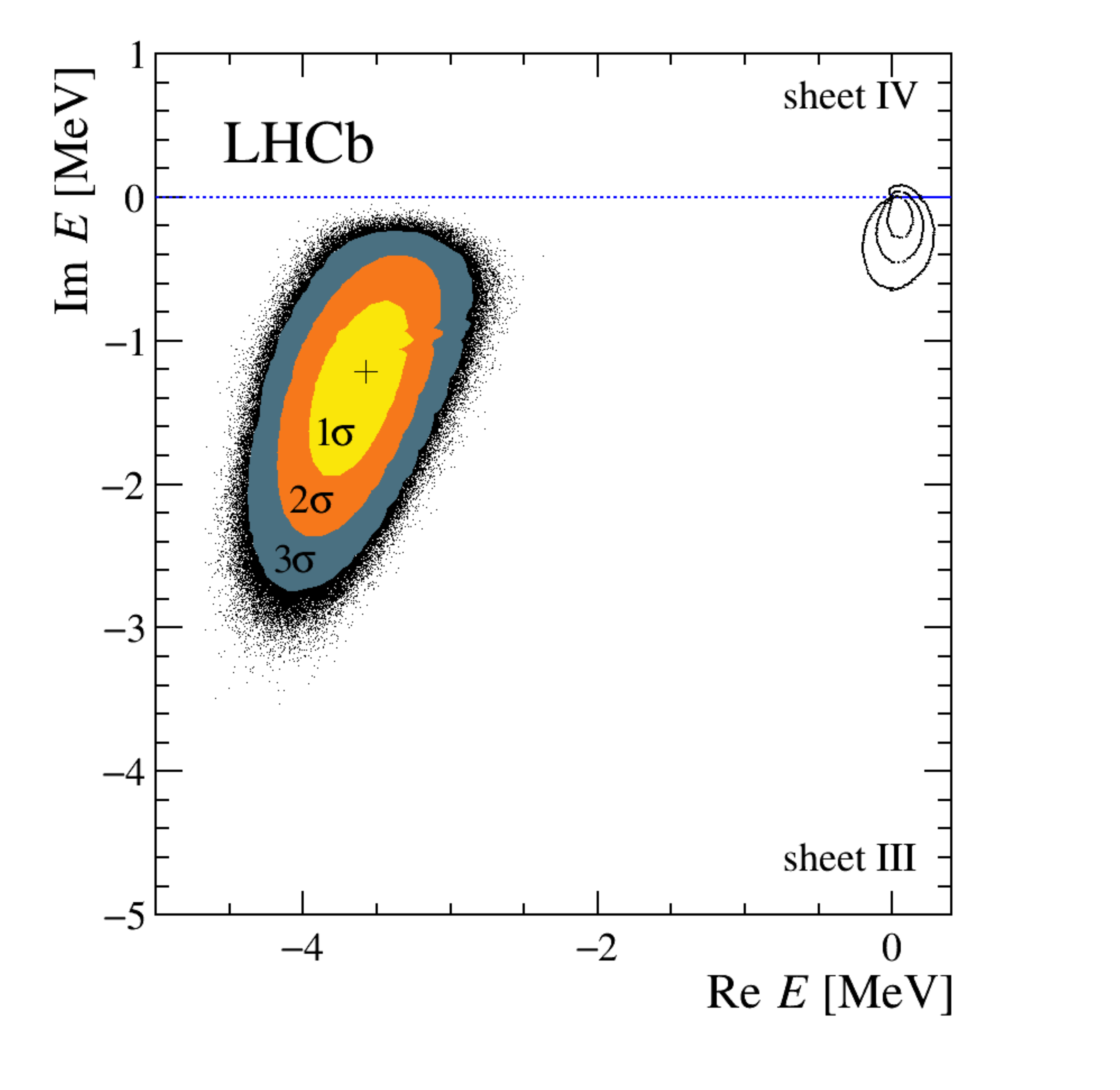}
 \caption{\small Confidence regions for the pole position
    on sheet\,III in the complex energy plane. The displayed uncertainties
   include statistical contributions and the modeling uncertainty.
   The poles are extracted at a Flatt\'e mass point of $m_0=3864.5\mev$.
   The shaded areas are the 1, 2 and 3$\sigma$ confidence regions.
   The branch cut is shown as the blue line.
       The location of the branch cut singularity is
    indicated with a vertical bar at $E=0+0\:i$.
    The best estimate for the pole positions is indicated by a cross.
    The confidence region for the pole
    on sheets\,II/IV is shown in outline for comparison.
    The black points indicate the samples from
    the pseudoexperiments procedure that lie outside the $3\sigma$ region.
  }
  \label{fig:polesIII}
 \end{figure}

As shown in Table~\ref{tab:flattesystematics},
taking into account the finite width of the \Dstarzb has a small effect
on the Flatt\'e parameters. However, the analytic structure of the amplitude
close to the threshold is changed such that in this case the branch cut
is located in the complex plane at
\mbox{$\operatorname{Im} E=-\Gamma_{\Dstarz}/2$}.
The phase of the amplitude for this case is shown
in Fig.~\ref{fig:polewithDstwidth}.
The displaced branch cut is highlighted in black.
The pole is found at  $E'_{\mathrm{II}} = (25 -140\:i)\kev$
in a similar location to the case without taking
into account the \Dstarzb width.
In particular, the most likely pole position is on sheet II,
the physical sheet with respect to the \Dz\Dstarzb system.
The location of the pole on sheet III is found
to be $E'_{\mathrm{III}}=(-3.59 -1.05\:i)\mev$,
similar to the fit that does not account for \Dstarzb width.

The uncertainties of the Flatt\'e parameters are
propagated to the pole position
by generating large sets of pseudoexperiments,
sampling from the asymmetric Gaussian uncertainties
that describe the statistical  and the
systematic uncertainties introduced through the resolution
and background parameterisation.
The systematic uncertainty on the pole position due to
the momentum scale, location of the threshold and the choice
of the Flatt\'e mass parameter are discussed in the following.

The confidence regions for the location
of the poles, corresponding to $68.3\%,95.4\%$ and
$99.7\%$ intervals, are shown in Figures ~\ref{fig:polesII} and \ref{fig:polesIII}.
For large values of $g$ the pole on sheet\,II moves to sheet\,IV,
which is analytically connected to the former
along the real axis above threshold.
Therefore, sheet\,II~(for $\operatorname{Im} E<0$) and
sheet IV~(for $\operatorname{Im} E >0$) are
shown together for this pole. While a pole location on sheet\,II
is preferred by the data, a location on sheet\,IV is still allowed at the $2\sigma$ level.
The pole on sheet\,III is located well below threshold and
comparatively deep in the complex plane and is shown in Fig.~\ref{fig:polesIII}. For comparison,
the location of the confidence region for the first pole on
sheets\,II and IV are also indicated on
sheet\,III.

\begin{figure}[t]
\centering
\includegraphics[width=1.0\columnwidth]{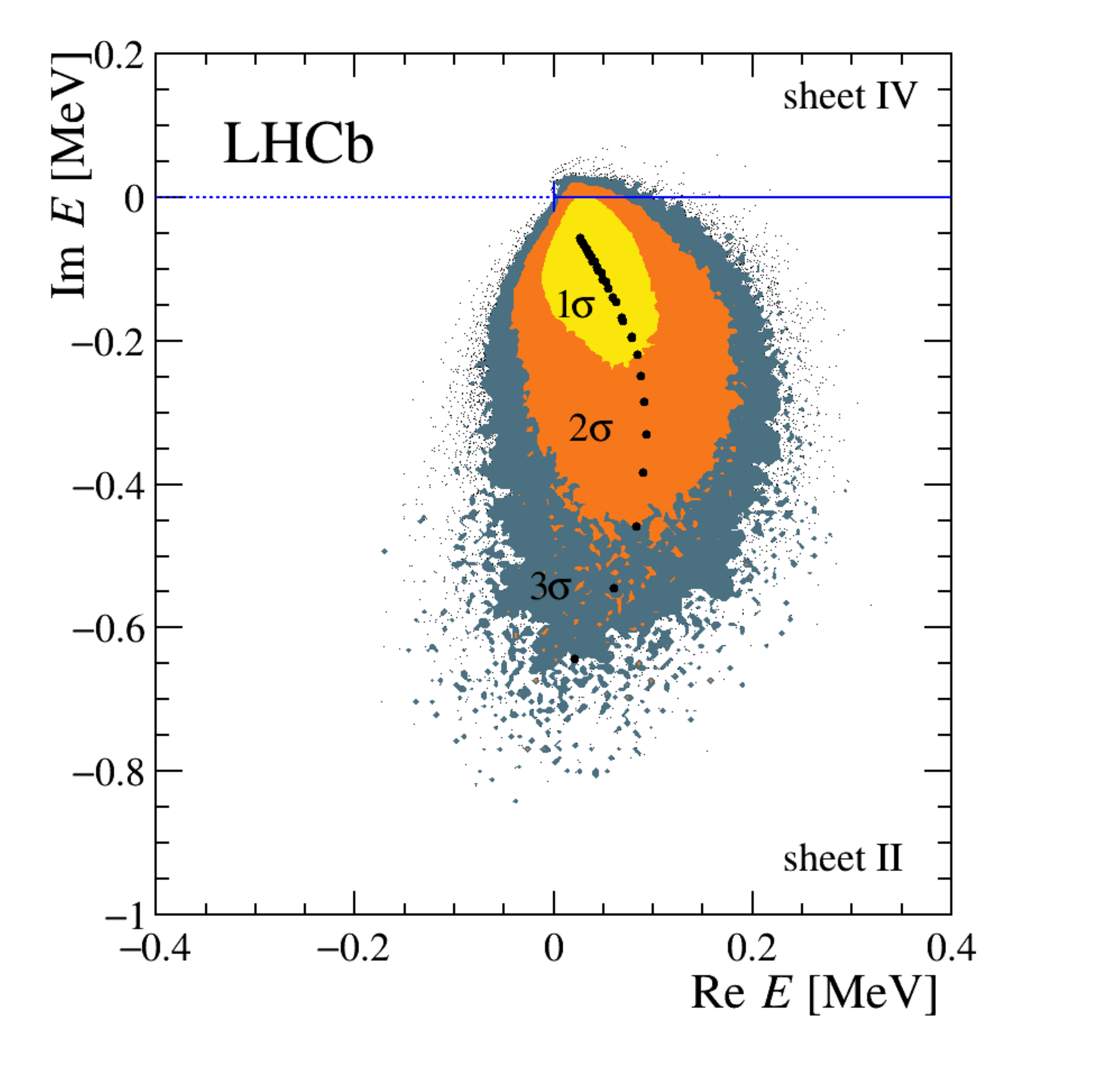}
\caption{\small Confidence regions for
  the pole on sheet\,II in the complex energy plane.
  The displayed uncertainties include statistical
  contributions and the uncertainty from the choice of
  the Flatt\'e mass parameter $m_0$.  Modelling
  uncertainties are not shown.
  The shaded areas are the 1, 2 and 3$\sigma$ confidence regions.
  The branch cut is shown as the blue line.
  The location of the branch cut singularity is
  indicated with a vertical bar at $E=0+0\:i$.
  The black circles indicate the best estimates for the
  pole position for the different choices of $m_0$.  }
\label{fig:pole_II_combined}
\end{figure}

The positions of both poles depend on the choice of
the Flatt\'e mass parameter $m_0$.
The dependence of the lineshape on $m_0$ has been explored in
the region below threshold and for the results shown in Fig.~\ref{fig:scanLL}
the corresponding pole positions are evaluated.
The location of the pole
on sheet II extracted for $-17 < E_f < 0\,\mev$
are marked by black circles in Fig.~\ref{fig:pole_II_combined}.
For smaller values of $m_0$ the pole moves closer to the real axis, for
values of $m_0$ approaching the threshold, the pole moves further
into the complex plane. For all fits performed the best estimate
for the location of the pole is on sheet II.

Figure~\ref{fig:pole_II_combined} also shows
the combined confidence regions, which account for
the explored range of $E_f$. For each fit, a sample
consisting of $10^5$~pseudoexperiments is drawn from
the Gaussian distribution described by the covariance
matrix of the fit parameters. Only the statistical
uncertainties obtained for each fit are used for this study.
The resulting samples of pole positions are combined by
weighting with their respective likelihood ratios with
respect to the best fit. The preferred location of
the pole is on sheet\,II. However, a location of
the pole on sheet\,IV is still allowed at the $2\sigma$ level.

The location of the pole on sheet\,III, in particular its real part,
depends strongly on the choice of $m_0$. For small values of $m_0$ the pole moves
away from the threshold and has less impact on the lineshape.
For $m_0$ approaching the threshold this pole moves closer
to the branch point and closer to the pole on sheet II.
Since the asymmetry of the poles with respect to
the threshold contains information on the potential
molecular nature of the state~\cite{poletrajectories},
the values of the pole positions are provided for
the most extreme scenario that is still allowed by
the data with a likelihood difference of
$\Delta\mathrm{LL}=1$,~($cf.$~Fig.~\ref{fig:scanLL})
at $m_0=3869.3\mev$.
In this case the two poles are found at
\mbox{$E_{\mathrm{II}}=(0.09-0.33\:i)\mev$} and
\mbox{$E_{\mathrm{III}}=(-0.85-0.97\:i)\mev$}.

\begin{figure}[tb]
\centering
\includegraphics[width=1.0\columnwidth]{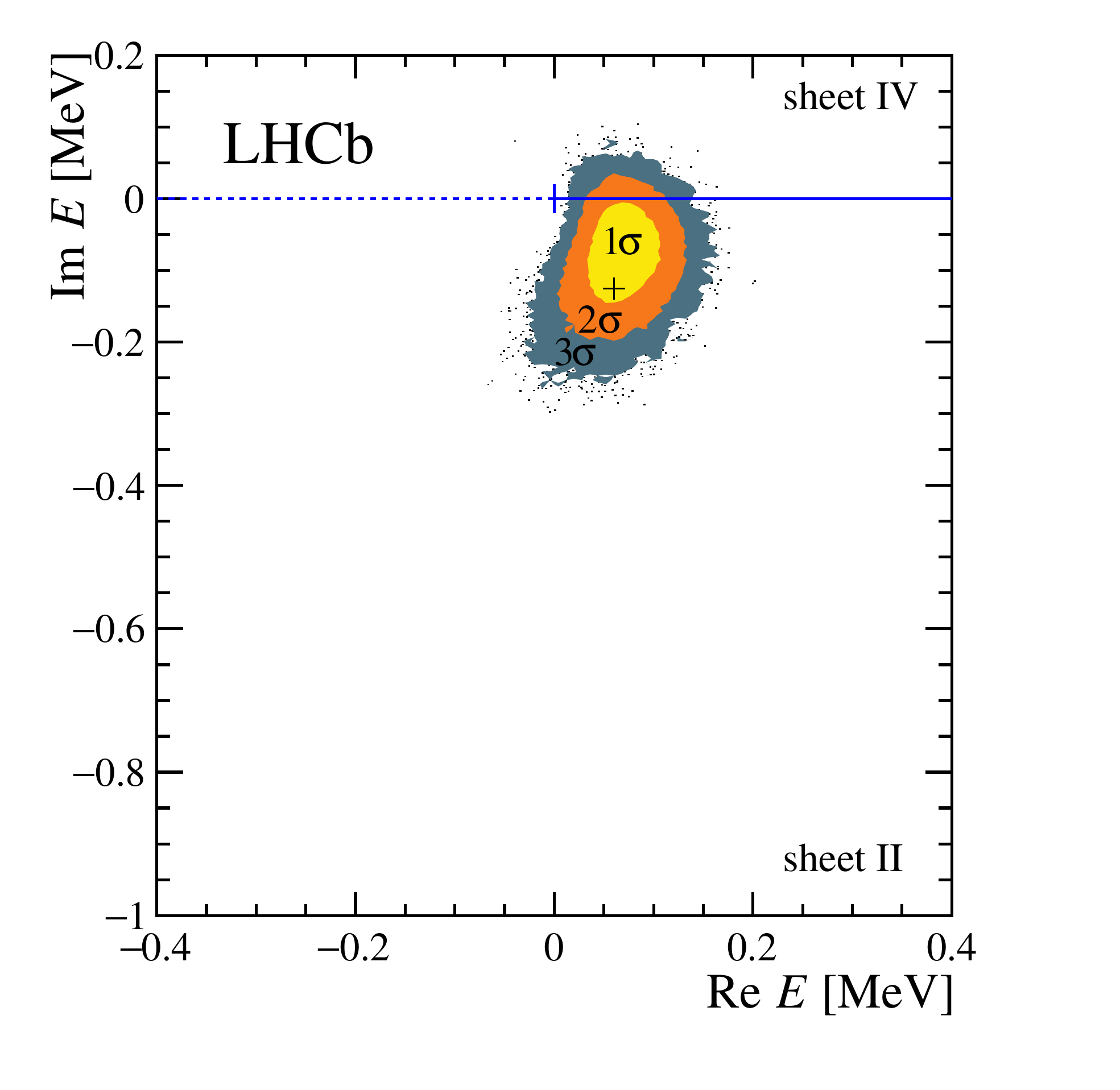}
  \caption{\small Confidence regions for the pole on
  sheet\,II in the complex energy plane, in the case
  that the mass scale is shifted up by $0.066\mev$,
  due to systematic uncertainty of the momentum scale.
  Only the statistical uncertainties are displayed.  The shaded areas are
  the 1, 2 and 3$\sigma$ confidence regions.
  The cross indicates the location of the pole found
  in the default fit, with the nominal momentum scale.
  The branch cut is shown as the blue line.
  The location of the branch cut singularity is
  indicated with a vertical bar at $E=0+0\:i$. }
\label{fig:pole_II_sfiftedmup}
\end{figure}

The location of the threshold with respect to the observed location
of the peak has a profound impact on the Flatt\'e parameters
and therefore on the pole position.
The main uncertainties, which
affect on which sheet the pole is found, are the
knowledge of the momentum scale and the location
of the $\Dz\Dstarzb$~threshold. As shown in
Table~\ref{tab:flattesystematics}, both effects are of equal importance.
Figure~\ref{fig:pole_II_sfiftedmup} shows the statistical uncertainties
of the pole on sheet\,II
for the case that the mass scale is shifted up
by $66\kev$. The pole is moving closer towards
the real axis but the preferred location remains on sheet\,II.
A measurement of the lineshape in the $\Dz\Dstarzb$ channel
is needed to further improve the knowledge on the impact
of the threshold location.

It is possible to study the behavior of the poles
in the limit where only the
$\D\Dstarb$~channels are considered.
The trajectory traced by the pole on sheet\,II when the
couplings to the other channels ($f_\rho$, $f_\omega$,
$\Gamma_0$) are sent to zero is indicated by
the red curve in Fig.~\ref{fig:flatteriemannresult}.
The coupling $g$ and the Flatt\'e mass parameter $E_f$
are kept fixed while taking this limit.
For the best fit solution the pole moves below
threshold and reaches the real axis at $E=-24\kev$
staying on the physical sheet with respect to the
$\Dz\Dstarzb$ threshold. This location is consistent
with a quasi-bound state in that channel with a
binding energy of $E_b=24\,\kev$.
If the pole lies in the allowed region on sheet\,IV,
taking the same limit also sends the pole onto the real
axis below threshold, but on the unphysical sheet with
respect to $\Dz\Dstarzb$. This situation corresponds to
a quasi-virtual state. Both types of solutions are
analytically connected along the real axis through
the branch cut singularity. Therefore, only upper
limits on the binding energy can be set. For the
bound state solution and only accounting for statistical uncertainties,
the result is $E_b<57\kev$ at 90\% confidence level\,(CL).
Including the systematic uncertainties
due to the choice of the model this limit
becomes $E_b<100\kev$ at $90\%$\,CL.
Setting the couplings to the other channels to zero
causes the pole on sheet\,III to move to the real axis as well,
reaching it at $E = -3.51\mev$. The corresponding values
extracted at the highest allowed value of $m_0=3869.3\mev$
are $E_b=29\kev$ for the bound state pole and $E_b=0.73\mev$
for the pole on the unphysical sheet.


\section{Results and discussion}
\label{sec:summary}
In this paper a
large sample of $\chicone(3872)$ mesons from
$\bquark$-hadron decays collected by LHCb in 2011 and 2012
is exploited to study the lineshape of the $\chicone(3872)$ meson. Describing
the lineshape with a Breit--Wigner function determines the mass splitting between the
$\chicone(3872)$ and $\psitwos$~states to be
\begin{equation*}
\Delta m = 185.598 \pm 0.067\pm 0.068\mev\,,
\end{equation*}
where the first uncertainty is statistical and the second  systematic.
Using
the known value of the \psitwos mass~\cite{PDG2019}
this corresponds to
\begin{equation*}
m_{\chicone(3872)} =
3871.695 \pm 0.067\pm 0.068\pm 0.010\mev \,,
\end{equation*}
where the third uncertainty is due to the knowledge of
the~\psitwos mass. The result is
in good agreement with the current world average~\cite{PDG2019}. The
uncertainty is improved by a factor of two compared to the best previous
measurement by the CDF collaboration~\cite{Aaltonen:2009vj}. The measured value
can also be compared to the threshold value,
$m_{\Dz} + m_{\Dstarz} =
3871.70 \pm  0.11 \mev $. 
The $\chicone(3872)$ mass, evaluated from
the mean of a fit assuming the Breit--Wigner lineshape, is
coincident with the $\Dz \Dstarzb$ threshold
within uncertainties, with $\delta E = 0.01\pm0.14\mev$.
A non-zero Breit-Wigner width of the $\chicone(3872)$~state is obtained with a value of
\begin{equation*}
\Gamma_{\mathrm{BW}} =  1.39 \pm 0.24 \pm 0.10 \mev.
\end{equation*}
The values found here for $m_{\chicone(3872)}$
and $\Gamma_{\mathrm{BW}}$ are in good agreement
with a complementary analysis using fully
reconstructed
$\Bu \rightarrow \chicone(3872) \Kp$ decays
presented in Ref.~\cite{LHCb-PAPER-2020-009} and combined therein.

Since
\mbox{$\left| \delta E \right| < \Gamma_{\mathrm{BW}}$},
the value of $\Gamma_{\mathrm{BW}}$
needs to be interpreted
with caution as coupled channel effects distort the lineshape. To
elucidate this, fits using the Flatt\'e parameterization discussed in
Refs.~\cite{PhysRevD.76.034007,Kalashnikova:2009gt} are
performed.
The parameters are found to be
\begin{eqnarray*}
 g         & = &  0.108\pm0.003^{\,+\,0.005}_{\,-\,0.006}  \,, \\
 f_{\Prho} & = & \left(1.8\pm0.6^{\,+\,0.7}_{\,-\,0.6}\right)\times 10^{-3}        \,, \\
 \Gamma_0  & = & 1.4\pm0.4\pm 0.6 \mev \,,
\end{eqnarray*}
with $m_0$ fixed at $3864.5\mev$.
The mode
of the Flatt\'e  distribution agrees with the mean of the Breit--Wigner lineshape.
However, the determined FWHM
is much
smaller,
\mbox{$0.22^{\,+\,0.06\,+\,0.25}_{\,-\,0.08\,-\,0.17}\,\mev$},
highlighting the importance of a physically well-motivated
lineshape parameterization.
The sensitivity of the data to the tails of the mass distribution
limits the extent to which the Flatt\'e parameters can be determined,
as is expected in the case of a strong coupling of the state
to the $\Dz\Dstarzb$ channel \cite{Baru2005}.
Values of the parameter $E_f$ above $-2.0\mev$ are
excluded at 90\% confidence level.
The
allowed region below threshold is $-270< E_f <-2.0\mev$.
In this region a linear dependence between the
parameters is observed. The slope $\frac{dg}{dE_f}$ is
related to the real part of the scattering
length~\cite{PhysRevD.76.034007} and is measured to be
\begin{equation*}
    \dfrac{dg}{dE_f} = \left(-15.11\pm0.16\right)\gev^{-1}\,.
\end{equation*}
In order to investigate the nature of
the $\chicone(3872)$ state, the analytic
structure of the amplitude in the vicinity
of the $\Dz\Dstarzb$ threshold is examined.
Using the Flatt\'e amplitude, two poles are found.
Both poles appear on unphysical sheets with
respect to the $\jpsi\pip\pim$ channel and
formally can be classified as resonances.
With respect to the $\Dz\Dstarzb$ channel,
one pole appears on the physical sheet,
the other on the unphysical sheet.
This configuration, corresponding to
a quasi-bound $\Dz\Dstarzb$ state,
is preferred for all scenarios studied in this paper.
However, within combined statistical and systematic
uncertainties a location of the first pole
on the unphysical sheet is still allowed
at the $2\sigma$ level and a quasi-virtual state
assignment for the $\chicone(3872)$ state cannot be excluded.

For the preferred quasi-bound state scenario the $90\%$ CL upper limit of
the $\Dz\Dstarzb$ binding energy $E_b$ is found to be $100\,\kev$.
The asymmetry of the locations of the two poles,
which is found to be substantial,
provides information on the composition of the $\chicone(3872)$ state.
In the case of a dominantly molecular nature of a state
a single pole close to  threshold is expected, while in
the case of a compact state there should be two
nearby poles~\cite{MORGAN1992632}. The argument is equivalent to
the Weinberg composition criterion~\cite{WeinbergComp} in the sense
that the asymmetry of the pole location in momentum space determines
the relative fractions of  molecular and  compact components
in the $\chicone(3872)$ wave function~\cite{BARU200453}
\begin{equation*}
    \dfrac{\left|k_2\right|-\left|k_1\right|}
    {\left|k_1\right|+\left|k_2\right|} = 1-Z\,.
\end{equation*}
Here $Z$ is the probability to find a compact component
in the wave function. The momentum $\left|k_1\right|=6.8\mev$
is obtained by inserting the binding energy of the bound state
pole into Eq.~\eqref{eqn:breakup}. The corresponding value
for the second pole is $\left|k_2\right|=82\mev$ and therefore
one obtains $Z=15\%$.
The asymmetry of the poles depends on the choice of $m_0$.
The asymmetry
is reduced as the $m_0$ parameter approaches the threshold.
The largest value for $m_0$ that is still compatible
with the data is $3869.3\mev$.
In this case one obtains $Z=33\%$ and therefore
the probability to find a compact component
in the $\chicone(3872)$ wave function is less than a third.
It should be noted that this argument depends on the
extrapolation to the single channel case. For resonances
the wave function normalisation used in the Weinberg criterion
is not valid and $Z$ has to be replaced by an integral over
the spectral density~\cite{BARU200453}. Nevertheless,
the value obtained in this work is in agreement with the
results of the analysis of the spectral density using
Belle data \cite{Adachi:2008sua,Adachi:2008te},
presented in Ref.~\cite{Kalashnikova:2009gt}.

The results for the amplitude parameters and in particular
the locations of the poles, are systematically
limited. In the future, a combined analysis
of the~\mbox{$\decay{\chicone(3872)}{\jpsi\pip\pim}$} and
\mbox{$\decay{\chicone(3872)}{\Dz\Dstarzb}$} channels
will make possible improvements
to the knowledge on the amplitude parameters.

\section*{Acknowledgements}
%
%
\noindent We thank C.~Hanhart and A.~Pilloni for useful discussions on the Flatt\'e model and the analytic structure of the amplitude.
We express our gratitude to our colleagues in the CERN
accelerator departments for the excellent performance of the LHC. We
thank the technical and administrative staff at the LHCb
institutes. We acknowledge support from CERN and from the national
agencies: CAPES, CNPq, FAPERJ and FINEP (Brazil); MOST and NSFC
(China); CNRS/IN2P3 (France); BMBF, DFG and MPG (Germany); INFN
(Italy); NWO (The Netherlands); MNiSW and NCN (Poland); MEN/IFA
(Romania); MinES and FASO (Russia); MinECo (Spain); SNSF and SER
(Switzerland); NASU (Ukraine); STFC (United Kingdom); NSF (USA).  We
acknowledge the computing resources that are provided by CERN, IN2P3
(France), KIT and DESY (Germany), INFN (Italy), SURF (The
Netherlands), PIC (Spain), GridPP (United Kingdom), RRCKI and Yandex
LLC (Russia), CSCS (Switzerland), IFIN-HH (Romania), CBPF (Brazil),
PL-GRID (Poland) and OSC (USA). We are indebted to the communities
behind the multiple open-source software packages on which we depend.
Individual groups or members have received support from AvH Foundation
(Germany), EPLANET, Marie Sk\l{}odowska-Curie Actions and ERC
(European Union), ANR, Labex P2IO and OCEVU, and R\'{e}gion
Auvergne-Rh\^{o}ne-Alpes (France), RFBR, RSF and Yandex LLC (Russia),
GVA, XuntaGal and GENCAT (Spain), Herchel Smith Fund, the Royal
Society, the English-Speaking Union and the Leverhulme Trust (United
Kingdom).


%


\clearpage
\addcontentsline{toc}{section}{References}
\setboolean{inbibliography}{true}
\bibliographystyle{LHCb}
\bibliography{local,LHCb-PAPER,standard,LHCb-CONF,LHCb-DP,LHCb-TDR}

\newpage


 
\newpage
\centerline
{\large\bf LHCb collaboration}
\begin
{flushleft}
\small
R.~Aaij$^{31}$,
C.~Abell{\'a}n~Beteta$^{49}$,
T.~Ackernley$^{59}$,
B.~Adeva$^{45}$,
M.~Adinolfi$^{53}$,
H.~Afsharnia$^{9}$,
C.A.~Aidala$^{82}$,
S.~Aiola$^{25}$,
Z.~Ajaltouni$^{9}$,
S.~Akar$^{64}$
% ...

{\footnotesize \it

% ...
$ ^{9}$Universit{\'e} Clermont Auvergne, CNRS/IN2P3, LPC, Clermont-Ferrand, France\\
% ...
$ ^{25}$INFN Sezione di Milano, Milano, Italy\\
% ...
$ ^{31}$Nikhef National Institute for Subatomic Physics, Amsterdam, Netherlands\\
% ...
$ ^{45}$Instituto Galego de F{\'\i}sica de Altas Enerx{\'\i}as (IGFAE), Universidade de Santiago de Compostela, Santiago de Compostela, Spain\\
% ...
$ ^{49}$Physik-Institut, Universit{\"a}t Z{\"u}rich, Z{\"u}rich, Switzerland\\
% ...
$ ^{53}$H.H. Wills Physics Laboratory, University of Bristol, Bristol, United Kingdom\\
% ...
$ ^{59}$Oliver Lodge Laboratory, University of Liverpool, Liverpool, United Kingdom\\
% ...
$ ^{64}$University of Cincinnati, Cincinnati, OH, United States\\
% ...
$ ^{67}$Syracuse University, Syracuse, NY, United States\\
% ...
$ ^{82}$University of Michigan, Ann Arbor, United States, associated to $^{67}$\\
\bigskip
% ...
\medskip
}
\end{flushleft}

\end{document}